\documentclass[preprint,aps,epsfig]{revtex4}
\draft

\usepackage{graphicx}% Include figure files
\usepackage{dcolumn}% Align table columns on decimal point
\usepackage{bm}% bold math
\usepackage{epsfig}

\begin{document}

\title{Superconductivity Versus Phase Separation, Stripes, and Checkerboard
Ordering: A Two-Dimensional Monte Carlo Study}
\author{D. Valdez-Balderas}
\email{balderas@mps.ohio-state.edu}
\author{David Stroud}
\email{stroud@mps.ohio-state.edu}
\address{Department of Physics, The Ohio State University, Columbus, Ohio 43210}
\date{\today}

\begin{abstract}

Using Monte Carlo techniques, we study a simple model which exhibits
a competition between superconductivity and other types of order in
two dimensions.  The model is a site-diluted XY model, in which the
XY spins are mobile, and also experience a repulsive interaction
extending to one, two, or many shells of neighbors.  Depending on
the strength and range of the repulsion and spin concentration, the
spins arrange themselves into a remarkable variety of patterns at
low temperatures $T$, including phase separation, checkerboard
order, and straight or labyrinthine patterns of stripes, which
sometimes show hints of nematic or smectic order.   This pattern
formation profoundly affects the superfluid density, $\gamma$. Phase
separation tends to enhance $\gamma$, checkerboard order suppresses
it, and stripe formation increases the component of $\gamma$ parallel
to the stripes and reduces the perpendicular one. We verify that
$\gamma(T = 0)$ is proportional to the effective conductance of a
random conductance network whose conductances equal the couplings of
the XY system. Possible connections between the model and real
materials, such as single high-T$_c$ cuprate layers, are briefly
discussed.
%DANIEL: note change of abstract, and mention of nematic and smectic order

\end{abstract}

\maketitle
%%%%%%%%%%%%%%%%%%%%%%%%%%%%%%%%%%%%%%%%%%%%%%%%%%%%%%%%%%%%%%%%%%%%%%%%%
\section{Introduction}

Systems with competing interactions have been known to show complex
patterns of self-organization~\cite{seul_andelman}. Examples of such
systems in two-dimensions include the following:
type I superconducting films in their intermediate
state~\cite{huebener}; adsorbed monolayers on surfaces~\cite{kern};
monomolecular film at air-water interfaces~\cite{seul_sammon};
ferrimagnetic garnet films~\cite{seul_wolfe}; and doped
%DANIEL: do you mean ferrimagnetic, or ferromagnetic?
%DR. STROUD, I verified it is ferrimagnetic garnet films
antiferromagnets, on which there are a number of reports of magnetic
and charge ordering~\cite{cheong_aeppli, yamada, tranquada, niemoller,
  mook, arai, pengcheng, hayden, sun}.
%DANIEL, can you give a more complete list than these?  Also, it isn't clear
%that what we are considering actually involves competing interactions.
%
% DR. STROUD, I added more references. In regards to the competing interactions,
% with the added section of a Coulomb repulsion, can we now say that  our
% system possesses competing interactions at low temperatures?
Corresponding to those experimental reports, there have been a number
of theoretical models intended to describe them.  For
example, numerous authors have studied Ising models that
include a short range ferromagnetic interaction plus a longer-ranged
antiferromagnetic interaction~\cite{booth, stoycheva, macisaac,
  low_emery, jamei, abanov, arlett, hurley, czeck, iglesias}. The
origin of those interactions vary, depending on the particular
system under study. In studies of magnetic films, for instance, the
origin of the ferromagnetic term is the exchange interaction between
spins, while the antiferromagnetic term has its origin in
dipole-dipole interactions.  In the context of hole-doped
antiferromagnets, charge stripes are
believed~\cite{emery_kivelson_tranquada} to originate from a
short-ranged tendency of the holes to accumulate in regions of
suppressed antiferromagnetism, frustrated by a long-ranged Coulomb
repulsion between them.  Other workers \cite{reichhardt} have treated
statics and dynamics of stripe phases in the presence of quenched
disorder.

Besides charge stripes, other forms of electronic inhomogeneities at
low temperatures have been observed in doped antiferromagnets. For
example, granular structures consisting of superconducting domains
separated by non-superconducting regions have been
reported~\cite{lang_davis} in scanning tunneling microscopy (STM)
studies of underdoped Bi$_{2}$Sr$_{2}$CaCu$_{2}$O$_{8+x}$; spatial
variations of the superconducting energy gap and of the local density
of states spectrum has been observed in STM
experiments~\cite{pan_davis} on optimally doped Bi$_{2}$Sr$_{2}$CaCu$_{2}$O$_{8+x}$; and
studies of lightly doped Ca$_{2-x}$Na$_{x}$CuO$_{2}$Cl$_{2}$ show that
electronic states within certain energy ranges show spatial
modulations in the form of checkerboards~\cite{hanaguri_davis}.

This diversity of electronic structures observed in doped
antiferromagnets have prompted us to study an XY model with annealed
disorder, where different kind of geometrical orders might occur, and
to try to determine the interplay between those orders and
superconductivity.  Specifically, we have carried out a Monte Carlo
study of a site-diluted, two-dimensional XY-model on a square lattice,
in which, besides the XY coupling, there is an additional interaction
between mobile spins.  We have considered two types of such additional
interactions: (i) a screened Coulomb repulsive interaction between
spins, and (ii) a repulsive interaction between either
nearest-neighbor spins or {\em second} nearest-neighbor spins.  We
assume that this Hamiltonian governs an {\em annealed} system, in the
sense that the spins are free to move under the influence of the
repulsive interaction.  They are also free to order in response to the
XY interaction.  XY models have been used to study systems such as
granular superconductors~\cite{stroud_zeng, stroud_shih}, high
temperature bulk superconductors~\cite{carlson_manousakis},
two-dimensional superfluids~\cite{schultka_manousakis}, and the
superfluid transition of helium in porous media~\cite{moon}. However,
to the best of our knowledge, there are no studies of XY models with
annealed disorder as the one we present here.

Our main purpose in the present work is to investigate (i) what are
the low temperature geometrical orders that occur in an XY model
with annealed disorder and different types of repulsion between
spins, and (ii) how this rearrangement of spins affects the helicity
modulus $\gamma$ of the system.
First, we calculate the XY transition temperature for various choices
of the parameters.  This transition appears to be of the
Kosterlitz-Thouless variety, though we have not carried out detailed
finite-size scaling tests of this hypothesis. We also calculate the
behavior of the helicity modulus $\gamma$, which behaves like a tensor
for some of the low-temperature  phases.  Finally, we
determine the temperatures at which other transitions occur.  These
other transitions are induced by the Ising-like spin-spin repulsion
mentioned above, and are of two types: (i) an "antiferromagnetic"
transition into a checkerboard-like structure for nearest-neighbor
repulsion, and (ii) a transition into a stripe phase for
second-nearest-neighbor repulsion.  In both cases, we find that the
transition has a strong effect on $\gamma$, which is described in
detail below.

Our calculations are carried out using standard Monte Carlo (MC)
simulations.  In some cases, we calculate critical values of the
parameters that determine the type of low-temperature structures. We
also compare our results to previous studies of site diluted XY-models
with {\em quenched} disorder~\cite{stroud_zeng}. The behavior in the
quenched and annealed cases are notably different, as discussed
further below.

The remainder of this paper is organized as follows.  In Section II,
we describe the studied models. Following this, we give a brief
description of our computational method in Section III.  Our results
are presented in Section IV, and a discussion of them is given in
Section V.

%%%%%%%%%%%%%%%%%%%%%%%%%%%%%%%%%%%%%%%%%%%%%%%%%%%%%%%%%%%%%%%%%%%%%%%%%
\section{Model}

We consider two distinct model Hamiltonians.  The first model, which
we denote Model I, consists of a two-dimensional (2D) square lattice
with a fraction $p$ of sites occupied by planar spins, and the
remaining $1 - p$ vacant. The spins have fixed length but are
characterized by an angle $\theta_i$, and are assumed to be
described by the following classical Hamiltonian:
\begin{equation}
  H = - \sum_{<i,j>} J_{ij} \cos(\theta_{i}-\theta_{j})\; + \;
  C\; \sum_{i \neq j}\;  n_{i} n_{j} \frac {\exp(-r_{ij}/r_{c}) }{r_{ij}}
  \label{eq:hamil1}
\end{equation}
Here $J_{ij}=J > 0$ if the sites $i$ and $j$ are both occupied, and
$J_{ij}=0$ otherwise.  The first sum is taken over distinct pairs of
nearest-neighbors $<i,j>$, while the second sum is taken over all
distinct pairs $(i,j)$ of lattice sites separated by a distance $r_{ij} <
r_{cut}$, where $r_{cut}$ is some cutoff radius, usually taken to be
a few times the screening length $r_c$. $C$ is a non-negative
constant that specifies the repulsion strength, and $n_i = 1$ or $0$
is the number of spins at site $i$. The first term thus corresponds
to the standard $XY$ Hamiltonian in 2D, while the
second term represents a screened Coulomb repulsion
between the spins, which is independent of the angles $\theta_i$ and
$\theta_j$. Thus, each site has two degrees of freedom: the spin
variable $\theta_i$, and the occupation number $n_i$.

% added to the first resubmission
Physically, each site which is occupied by a spin can be interpreted
as a mobile, charged superconducting domain, and each vacant site as a
negatively charged non-superconducting region.  The angles $\theta_i$
of the ``spins'' represent the phases of the superconducting order
parameters in the i$^{th}$ domain.  When two superconducting domains
are close to each other, they couple via Josephson tunneling, which is
the origin of the first term in Eq.~(\ref{eq:hamil1}).

The second term in Eq.~(\ref{eq:hamil1}) originates in the Coulomb repulsion
between charged superconducting domains.   While this term  might
appear to violate overall charge neutrality, it is, in fact,
consistent with that requirement, as we now show.  The screened
Coulomb term [the second term in Eq.~(\ref{eq:hamil1})] should really be written as
\begin{eqnarray}
H_{Coul} = \sum_{i\neq j}q_iq_j\frac{\exp(-r_{ij}/r_c)}{r_{ij}},
\nonumber
\end{eqnarray}
where the sums run over all pairs of lattice sites, and charge
neutrality requires that $\sum_iq_i = 0$.  We assume that $q_i =
q_S$ or $q_I$ on the superconducting or insulating sites, where
$pq_S + (1 - p)q_I = 0$.%, and $p$ is the fraction of superconducting sites.
Now note that
\begin{eqnarray}
H_{Coul}  = \sum_{i\neq j}
(q_i-q_I)(q_j-q_I)\frac{\exp(-r_{ij}/r_c)}{r_{ij}} +q_I\sum_{i\neq
j}(q_i + q_j)\frac{\exp(-r_{ij}/r_c)}{r_{ij}}- \nonumber \\ -
q_I^2\sum_{i\neq j}\frac{\exp(-r_{ij})/r_c}{r_{ij}}.\nonumber
\end{eqnarray}
But in this last equation, the second term vanishes because
$\sum_iq_i=0$, while the third is just an additive constant. The
summand of the first term is non-vanishing only on the $S$ sites.
Thus, $H_{Coul}$ can be rewritten, to within a constant, as
\begin{eqnarray}
H_{Coul} = \sum_{i\neq j}n_in_j(q_S -
q_I)^2\frac{\exp(-r_{ij}/r_c)}{r_{ij}}.\nonumber
\end{eqnarray}
This form is indeed equivalent to the second term in
Eq.~(\ref{eq:hamil1}) with $C = (q_S - q_I)^2$.

At high temperatures $T$, the system described by
Eq.~(\ref{eq:hamil1}) is expected to be phase-incoherent, and the
spins will point in random directions.  As $T$ is reduced, neighboring
spins tend to align with each other in order to minimize the system
energy.  This short-ranged attraction between spins competes with the
longer-ranged screened Coulomb repulsion represented by the second
term in the Hamiltonian Eq.~(\ref{eq:hamil1}).  Thus, this Hamiltonian
is a simple model of a system with competing interactions.

We have also studied a model Hamiltonian ("Model II") consisting of
an XY term plus a short-range repulsion:
\begin{equation}
  H = - \sum_{<i,j>} J_{ij} \cos(\theta_{i}-\theta_{j}) +
  A \sum_{<i,j>}  n_{i} n_{j} + B \sum_{<i,j>'}  n_{i} n_{j}.
  \label{eq:hamil2}
\end{equation}
Here the first and second sums are carried out over distinct pairs
of nearest-neighbors $\langle i,j\rangle$, while the third sum runs over all
distinct pairs of second-nearest-neighbors.  $A$ and $B$ are
non-negative constants specifying the strength of the
nearest-neighbor and second-nearest-neighbor repulsion. As will be
shown, Hamiltonian~(\ref{eq:hamil2}) at low $T$ produces patterns
similar to those of Model I.

We obtain the equilibrium thermodynamics of these two models by
treating (\ref{eq:hamil1}) and (\ref{eq:hamil2}) classically. At a
given $T$, the $\theta_i$'s and $n_i$'s arrange themselves so as to
minimize the Helmholtz free energy $F$, subject to the constraint
that $\sum_i \langle n_i \rangle/N = p$, where $p$ is assumed to be
specified by the experimental conditions. Since both $\theta_i$ and
$n_i$ are free to change, these Hamiltonians describe systems with
annealed disorder. By contrast, in a typical system with quenched
disorder, the $n_i$'s are assumed fixed, and only the $\theta_i$'s
are free to change.  In this quenched case, the last two terms in
the Hamiltonian would have the same value for any configuration of
the XY spins, and thus play no role in determining the system
thermodynamics.

 %%%%%%%%%%%%%%%%%%%%%%%%%%%%%%%%%%%%%%%%%%%%%%%%%%%%%%%%%%%%%%%%%%%%%%%%%
\section{Computational Method}

\subsection{Monte Carlo Approach}

We have studied Models I and II for several different values of the
parameters $A$, $B$, and $C$, and for different spin concentrations
$p$, using a Monte Carlo (MC) approach.  We used a square lattice,
generally of size $30\times30$ sites, with periodic boundary
conditions, using the standard MC Metropolis
algorithm~\cite{barkema}.
%DANIEL, need a reference here and a little more description of the method.
% DR. STROUD, I added the reference above, and more description of the methods below.
In all of the studied systems we have set $J_{ij}=1$ if both sites
$i$ and $j$ are occupied, and $J_{ij}=0$ otherwise. For Model I, we
considered annealed systems with $ C \geq 0$, while for Model II, we
studied two classes of parameters: (i) a nearest-neighbor repulsion
only ($A \geq 0$; $B=0$), and (ii) second-nearest-neighbor repulsion
only ($A = 0$; $B \geq 0$).

We have also reproduced some previous MC studies of 2D systems with
quenched disorder~\cite{stroud_zeng}, in order to compare with our
present annealed results.  For the quenched systems (in which the
spins are distributed randomly at fixed locations on the lattice),
we averaged over 20 different quenched disorder realizations, each
with a spin concentration $p$.
%DANIEL, is this what you did?
% DR. STROUD, yes
For each disorder realization, we started the MC run with the spins
arranged in a random configuration of $\{\theta_i\}$'s and the
system at $T = 1.2$ (in units of $J/k_B$). The system was then
cooled in steps of $\Delta T=0.05$, down to $T = 0.1$, and $\Delta T
= 0.025$ for $T \leq 0.1$.  For each $T$, we carried out $5 \times
10^4$ sweeps through the entire lattice, taking averages over the
last $2 \times 10^4$ sweeps, where each sweep consisted of an MC
attempt to vary the angle of each spin.
% for the case spin concentration
% $p=0.9$of $p=0.8$
% and $p=0.7$, taking averages over the last 10000 sweeps. The cases of
% $p<=p_{c}(=0.59)$, where $p_{c}$ is the critical percolation
% threshold, the system with quenched disorder doesn't undergo a phase
% transition, so we haven't studied such systems with quenched
% disorder.

For systems with annealed disorder, we started the system with the
spins in randomly chosen sites on the lattice, with randomly chosen
angles $\theta_i$, at a starting $T = \text {max}[2 C,1.2]$ (here $C$ is
in units of $J/k_{B}$) for Hamiltonian~(\ref{eq:hamil1}), and $T=1.2$ for
Hamiltonian~(\ref{eq:hamil2}). The system was then cooled down to a
$T=0.025$, in steps of $\Delta T$.  For Model I, we took $\Delta T =
0.1 C$ for $T > 1.0$, $\Delta T = 0.05$ for $0.1 < T < 1$, and $\Delta
T = 0.025$ for $T < 0.1$. For Model II, we used the same $\Delta T$'s
as in the quenched case.  For each $T$, we carried out a series of MC
sweeps over both spin angles and spin positions. During a sweep over
the spin angles, MC changes in the spin angles were attempted, while
on sweeps over spin positions, the MC step consisted of an attempt to
move each of the vacancies to a randomly chosen occupied
site~\cite{barkema}. After each sweep over spin position, ten sweeps
over spin angles were carried out.  To find the low-$T$ spin
configuration of Model I, we took total of $2\times 10^6$ sweeps over
spin angles.  But to compute thermal averages for Model II, we carried
out $1\times 10^6$ sweeps over spin angles, discarding the first
$5\times 10^5$ in order to allow the system to equilibrate.  Following
this relaxation, the MC thermal averages were carried out every tenth
sweep over spin angles, immediately before a sweep over spin positions
was carried out.

For Model I, the second sum was taken only over those distinct pairs
of lattice sites which are separated by $r_{ij} < \text {min}[3
r_{c},L/2]$, where $L$ is the linear dimension of the square MC
cell.  [Thus, the cutoff radius $r_{cut}$ discussed below eq.\ (1)
is the lesser of $3r_c$ and $L/2$.]
%DANIEL2: Is the last sentence correct?
% DR STROUD: Yes.
This choice of $r_{ij} < L/2$ is necessary to avoid double
counting of pairs in when periodic boundary conditions are used, as
in the present calculation.

For both Model I and Model II, we have calculated several
equilibrium quantities.  To characterize phase coherence, we
computed the diagonal elements $\gamma_{\alpha \alpha}$
($\alpha=x,y$) of the helicity modulus tensor $\gamma$.  These
diagonal elements are the spin-wave stiffness constants of the XY
spin system, and, in a superconductor, are proportional to elements
of the superfluid density tensor.  They are defined as appropriate
equilibrium averages over the spin configuration.  $\gamma_{xx}$,
for example, is defined~\cite{stroud_zeng} by
%DANIEL2: I added several sentences above.
\begin{eqnarray}
 \gamma_{xx} =
 \frac{1}{N}  \langle \sum_{<i,j>} (x_i-x_j)^2 J_{ij} \cos(\theta_{i}-\theta_{j}) \rangle -
 \frac{1}{N k_{B} T} \langle  [ \sum_{<i,j>} (x_i-x_j) J_{ij} \sin(\theta_{i}-\theta_{j})]^{2} \rangle  \nonumber \\
 + \frac{1}{N k_{B} T} \langle  \sum_{<i,j>}(x_i-x_j) J_{ij} \sin(\theta_{i}-\theta_{j}) \rangle
 ^{2},
\label{helicity}
 \end{eqnarray}
 where $x_i$ is the $x$ coordinate of the spin on the lattice site
 $i$, $N$ is the total number of lattice sites,
 %DANIEL, whether occupied or not?
 % DR. STROUD, yes
 and $\langle \rangle$ denotes a canonical average.  $\gamma_{yy}$ is defined by the
 analogous expression with $x_i$ replaced by $y_i$.
%DANIEL, are you assuming that the lattice constant is unity?
 % DR. STROUD, yes
 In our computations, we have taken the lattice constant $a = 1$.
 To characterize the spins patterns, we have also calculated
 \begin{equation}
   S_{1}(\vec q, T) = <|n(\vec q)|^{2}>,
   \label{eq:SRO1}
 \end{equation}
and
 \begin{equation}
   S_{2}(\vec q, T) = <|n(\vec q)|^{2}>-|<n(\vec q)>|^{2},
   \label{eq:SRO2}
 \end{equation}
where
\begin{equation}
  n(\vec q) = \sum_{j=1}^{N} n_{j} \exp(-i \vec{q}\cdot\vec r_{j} \cdot
  ),
  \label{eq:nq}
\end{equation}
where $\vec r_{j}$ is the position of the $j$th lattice site. For
$\vec q=\pi \hat x / a$ and $\vec q=\pi \hat y / a$, $S_{1}(\vec q,
T)$ and $S_{2}(\vec q, T)$ probe stripe formation in the $y$ and $x$
directions, respectively; for $\vec q=\pi (\hat x +\hat y) / a$,
they probe the formation of checkerboard patterns, and for $\vec q=
2 \pi \hat x / (N_{x} a)$ and $\vec q= 2 \pi \hat y / (N_{y} a)$
(with $N_{x}$ and $N_{y}$ the number of lattice sites in the $x$ and
$y$-directions), they are sensitive to phase separation of the
system into large domains of occupied and vacant sites.

\subsection{Special Approach for Low-Temperature Helicity Modulus}

At low $T$, $\gamma_{xx}(T)$ and $\gamma_{yy}(T)$ can also be
calculated by using a mapping between these quantities and the
effective conductances $g_{e,xx}$ and $g_{e,yy}$ of a related
conductance network\cite{ebner_stroud}. Specifically, the helicity
modulus $\gamma_{\alpha\alpha}(T\rightarrow 0)$ in the $\alpha$
direction satisfies the following relation:
\begin{equation}
%  \frac{\gamma_{\alpha\alpha}(T=0, p)}{\gamma_{\alpha\alpha}(T=0,p = 1)} =
%  \frac{g_{e,\alpha\alpha}(p)}{g_{e,\alpha\alpha}( p =
%  1)}.
  \frac{\gamma_{\alpha\alpha}(T=0, p)}{\gamma_{\alpha\alpha}(T=0,p = 1)} =
  \frac{g_{e,\alpha\alpha}(p)}{g_{e,\alpha\alpha}( p =  1)}.
  \label{eq:conductivity}
\end{equation}
%
% DR. STROUD, as you know, in the systems with annealed disorder, $\gamma$ depends not only
% on the spin concentration, but also on the arrangement of the spins.
% Shouldn't we then use 'c' to denote a spin configuration instead of a spin
% concentration, and c_{0} instead of c=1, in order to denote the case of a system
% with no dilution?
%
Here $\gamma_{\alpha\alpha}(T=0,p)$ denotes the $\alpha,\alpha$
component of the helicity modulus tensor at $T = 0$ for a diluted
arrangement of XY spins of concentration $p$ and the actual $T = 0$
configuration, and $\gamma_{\alpha\alpha}(T = 0, p = 1)$ is the
(isotropic) helicity modulus of the corresponding array at $p = 1$.
$g_{e,\alpha\alpha}(p)$ and $g_{e,\alpha\alpha}(p = 1)$ are the
conductances of a certain conductance network associated with the
original array of diluted XY spins, and is constructed as follows:
We associate with each spin an electrical node, and with each
coupling constant $J_{ij}$ connecting spins at sites $i$ and $j$
we associate a "conductance" $g_{ij} = J_{ij}$. The effective
conductance of this network in the $\alpha$ direction ($\alpha = x$
or $y$) is denoted  $g_{e,\alpha\alpha}$. Eq.~(\ref{eq:conductivity})
allows us to calculate the ratios of helicity moduli at two
different concentrations $p$ from the corresponding conductances.
Our method of calculating the $g_e$'s needed for this mapping is
explained below.

 %%%%%%%%%%%%%%%%%%%%%%%%%%%%%%%%%%%%%%%%%%%%%%%%%%%%%%%%%%%%%%%%%%%%%%%%%
\section{Results and Discussion}

We have carried out extensive simulations for Models I and II,
considering systems of lattice size 30$\times$30, and a range of
$A$, $B$, $C$, $r_{c}$, $r_{cut}$, and $p$, as described below.
%DANIEL2: I added r_{cut} to the above list.
We first consider the low-$T$ ($T = 0.025$) spin configurations for
systems described by Model I.  Next, we present a more detailed
study of Model II results over a range of $T$.  While the two models
differ in how the repulsive interaction is truncated, several
features of the low-$T$ spin configurations prove to be
model-independent.

%%%%%%%%%%%%%%%%%%%%%%%%%%%%%%%%%%%%%%%%%%%%%%%%%%%%%%%%%%%%%%%%%%%%%%%%%

\subsection{Model I}

\begin{figure}
  \setlength{\unitlength}{1.0in}
  \begin{picture}(12,7)
    \put(0,0){
      \includegraphics*[width=5.5in, angle=0]{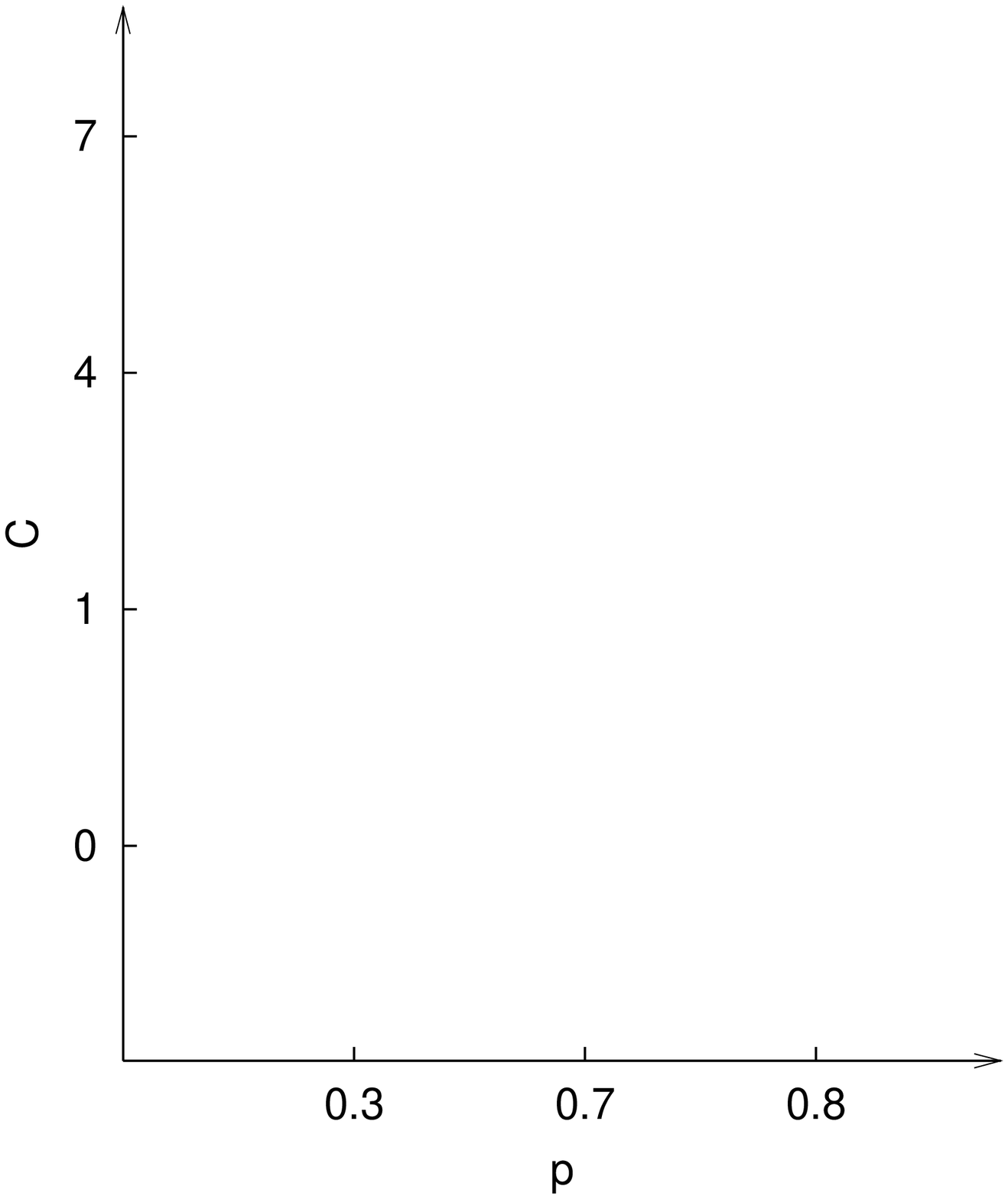}
    }

    \put(1.1,1){
      \includegraphics*[width=3.8in, angle=0]{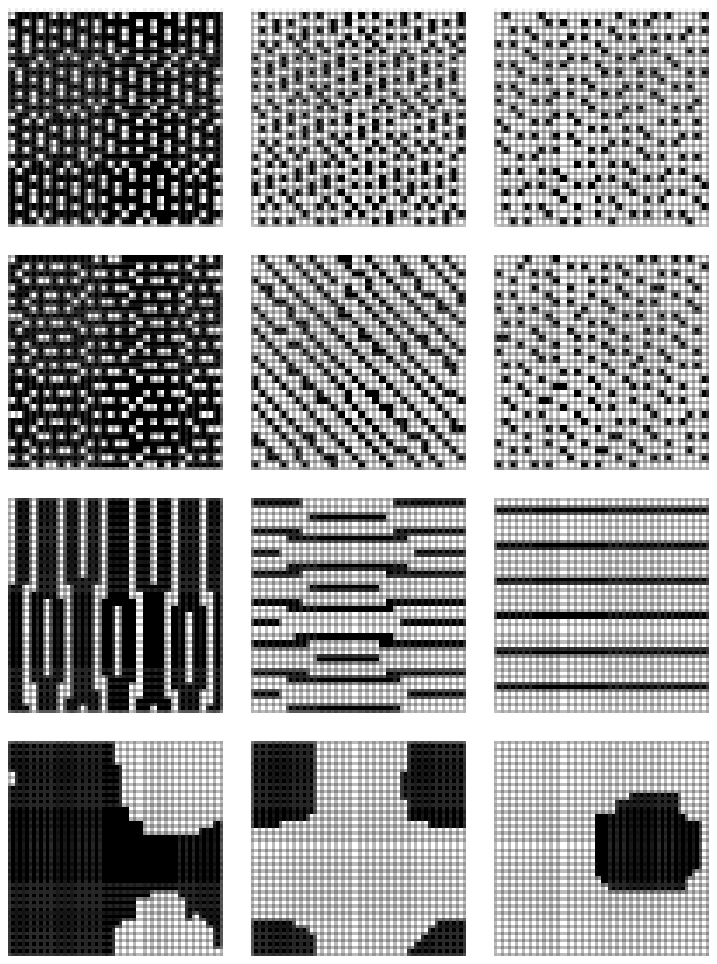}
    }
  \end{picture}
  \caption{\label{fig:pd_unscreened}
    Representative snapshots of the spin configurations at low temperatures (\mbox{$T = 0.025$})
    for model I at various points in the
    \mbox{$p$}-\mbox{$C$} phase diagram.  We show results for a system with an
    unscreened Coulomb repulsion truncated at $r = 15a$. $C$ is the
    strength of the Coulomb repulsion and $p$ is the spin
    concentration. A white (black) square is an occupied (vacant)
    site. Each snapshot was obtained by annealing the system from a
    $T = \text {max}[2 C,1.2]$ as described in the text.}
\end{figure}

%--------------COULOMB UNSCREENED
We have studied Model I [Hamiltonian~(\ref{eq:hamil1})] for several
values of the screening radius $r_c$. Low-$T$ configurations were
obtained by annealing the system starting from $T = \text {max}[2
C,1.2]$ to $T = 0.025$, as explained in the previous section.
Fig.~\ref{fig:pd_unscreened} shows the $T = 0.025$ spin configurations
for $r_{c}=\infty$.  This corresponds to a {\it unscreened} Coulomb
repulsion which is, however, truncated at $r = L/2$.  In this and all
later Figures, the white and black squares in the lattice correspond
to occupied and vacant sites.  At all $p$, in the absence of a Coulomb
repulsion ($C = 0$), the system phase-separates into regions of
occupied and vacant sites.  This behavior is expected, since this
configuration maximizes the number of nearest-neighbors for each spin,
and hence minimizes the internal energy.

At $C = 1$, phase separation no longer occurs; instead, the spins
arrange themselves into long, unidirectional stripes, whose width
increases with increasing spin concentration $p$.   For $C = 1$ and
$ p = 0.3$ and $0.7$, we see a kind of {\em smectic} pattern in the
stripes: the stripes seem to be arranged into layers which have a
characteristic thickness, though the spacing between stripes is not
perfectly ordered.  Such smectic states have been postulated (in the
context of a {\em quantum} model) for two-dimensional metallic
stripe phases in doped Mott insulators~\cite{emery}.  This spin
arrangement is a compromise between the clustering induced by the
short-range attractive XY interaction and the long-range repulsion
produced by the Coulomb interaction. As $C$ is increased further,
the system undergoes a characteristic series of morphology changes,
from long stripes, to shorter stripe-like patterns, and eventually
to a checkerboard pattern.  At some values of $p$ and large $C$,
there are suggestions that the occupied or vacant sites have a {\em
nematic} order - that is, they are arranged in short stripe-like
patterns with a preferred direction.  At certain values of $p$ and
$C$, we also see a pattern of {\em diagonal} stripes (at $p = 0.7$,
$C = 4$ in Fig.\ 1).  The checkerboard patterns are the state of
minimum energy when the Coulomb repulsion is much stronger than the
XY attraction.
%DANIEL2: I changed the above discussion - in particular, I added comments
%about "smectic order" - layers of parallel stripes.

\begin{figure}
  \setlength{\unitlength}{1.0in}
  \begin{picture}(12,7)
    \put(0,0){
      \includegraphics*[width=5.8in, angle=0]{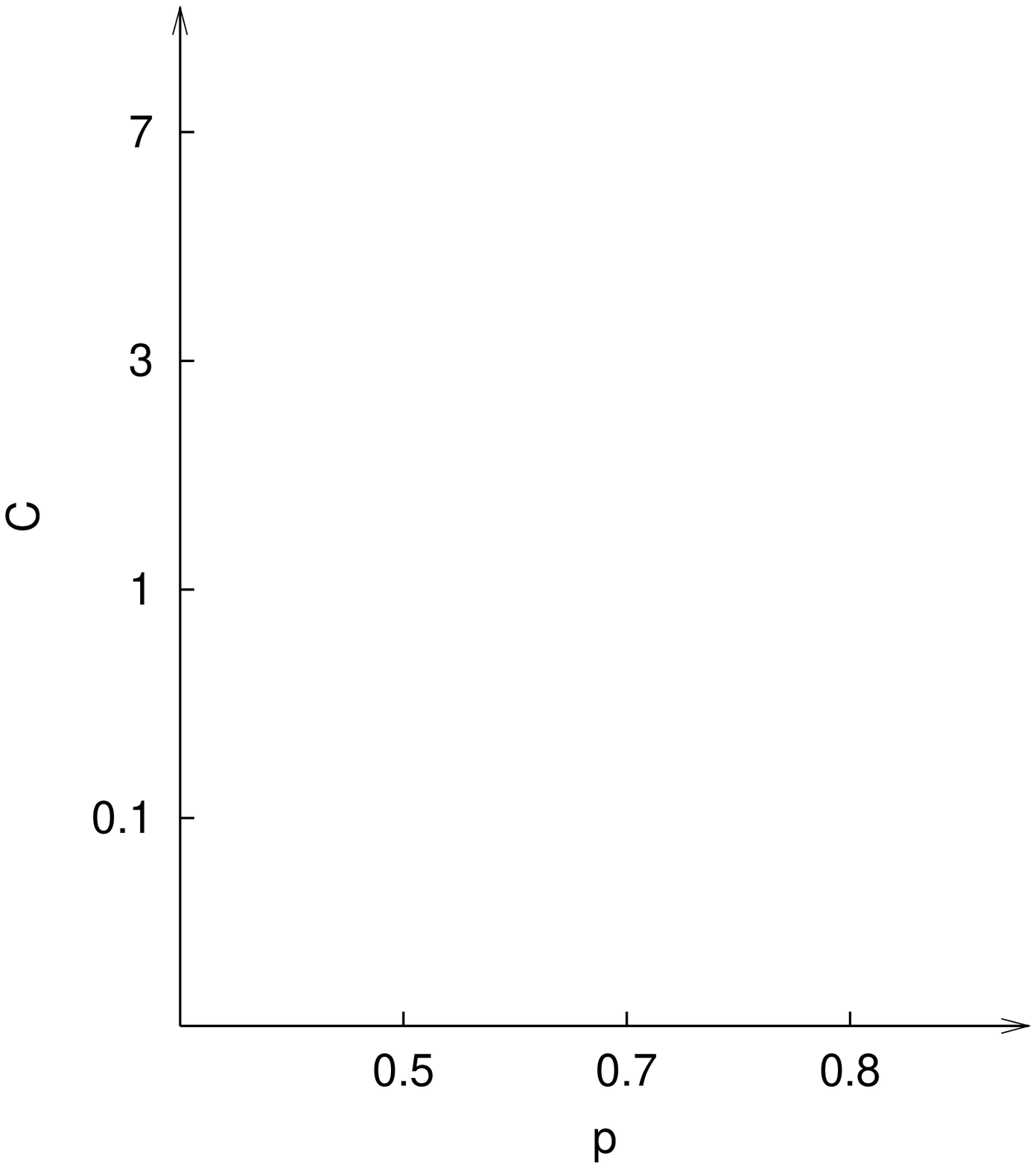}
    }

    \put(1.4,1.0){
      \includegraphics*[width=3.8in, angle=0]{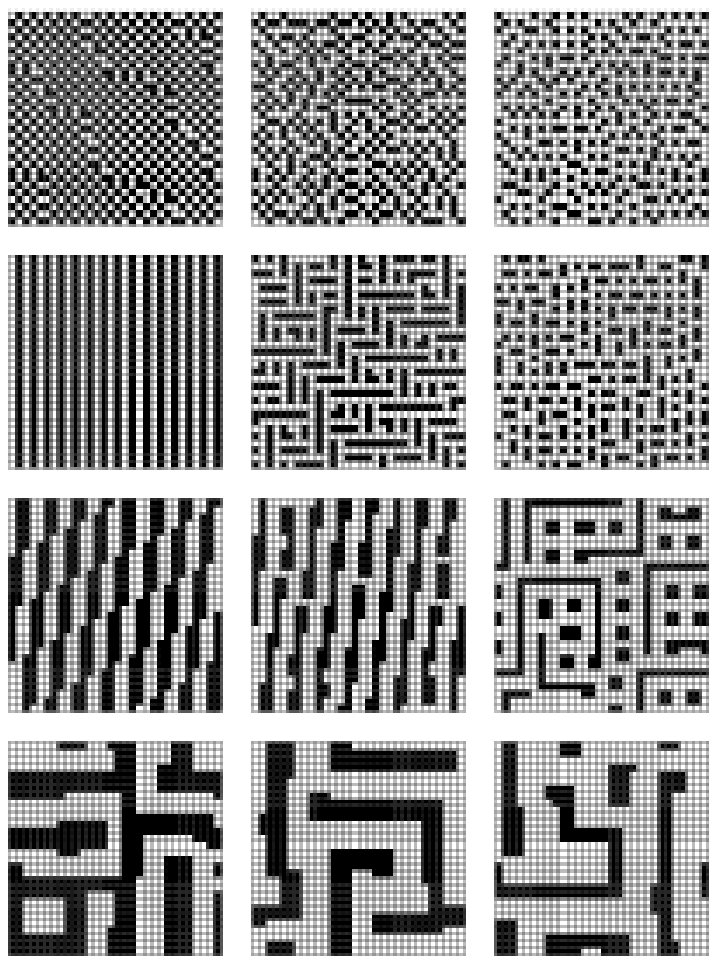}
    }
  \end{picture}
  \caption{\label{fig:pd_screened_rc7}
    Same as Fig.\ 1, but for a screened Coulomb repulsion with $r_c
    = 7a$.}
%    Sample of the low-temperature ($T = 0.025$) $p$-$C$ phase diagram
%    for a system with a screened Coulomb repulsion with $r_{c} = 7a$.}
\end{figure}

%----------- COULOMB SCREENED r_c = 7
We have also sampled the low-$T$ $p$-$C$ phase diagram of Model I
for finite $r_c$ (screened Coulomb repulsion).
Fig.~\ref{fig:pd_screened_rc7} shows snapshots of the spin
configurations for $r_c = 7$ and $p = 0.5$, $0.6$ and $0.8$.  The
destruction of the phase-separated case at $C = 0$ (shown only in
Fig.~\ref{fig:pd_unscreened}) by the screened Coulomb repulsion
proceeds first by formation of long, elongated domains of the
minority component (which, for the cases shown in
Fig.~\ref{fig:pd_screened_rc7}, are vacancies). These domains can be
seen in Fig.~\ref{fig:pd_screened_rc7} for $C = 0.1$. As $C$ is
increased, those elongated domains become unidirectional stripes [as
in ($p = 0.5, C = 1.0$), ($p = 0.5, C = 3.0$), and ($p = 0.7, C =
1.0$)], or long but tortuous stripes which coexist with small blobs
[($p = 0.8, A = 1.0$)]. As in Fig.~\ref{fig:pd_unscreened}, a
further increase in $C$ causes the stripes to break up and leads,
for the largest values of $C$, to checkerboard patterns, of which a
clear example is ($p = 0.5, A = 7.0$) in
Fig.~\ref{fig:pd_screened_rc7}.

\begin{figure}
  \setlength{\unitlength}{1.0in}
  \begin{picture}(12,7)
    \put(0,0){
      \includegraphics*[width=5.8in, angle=0]{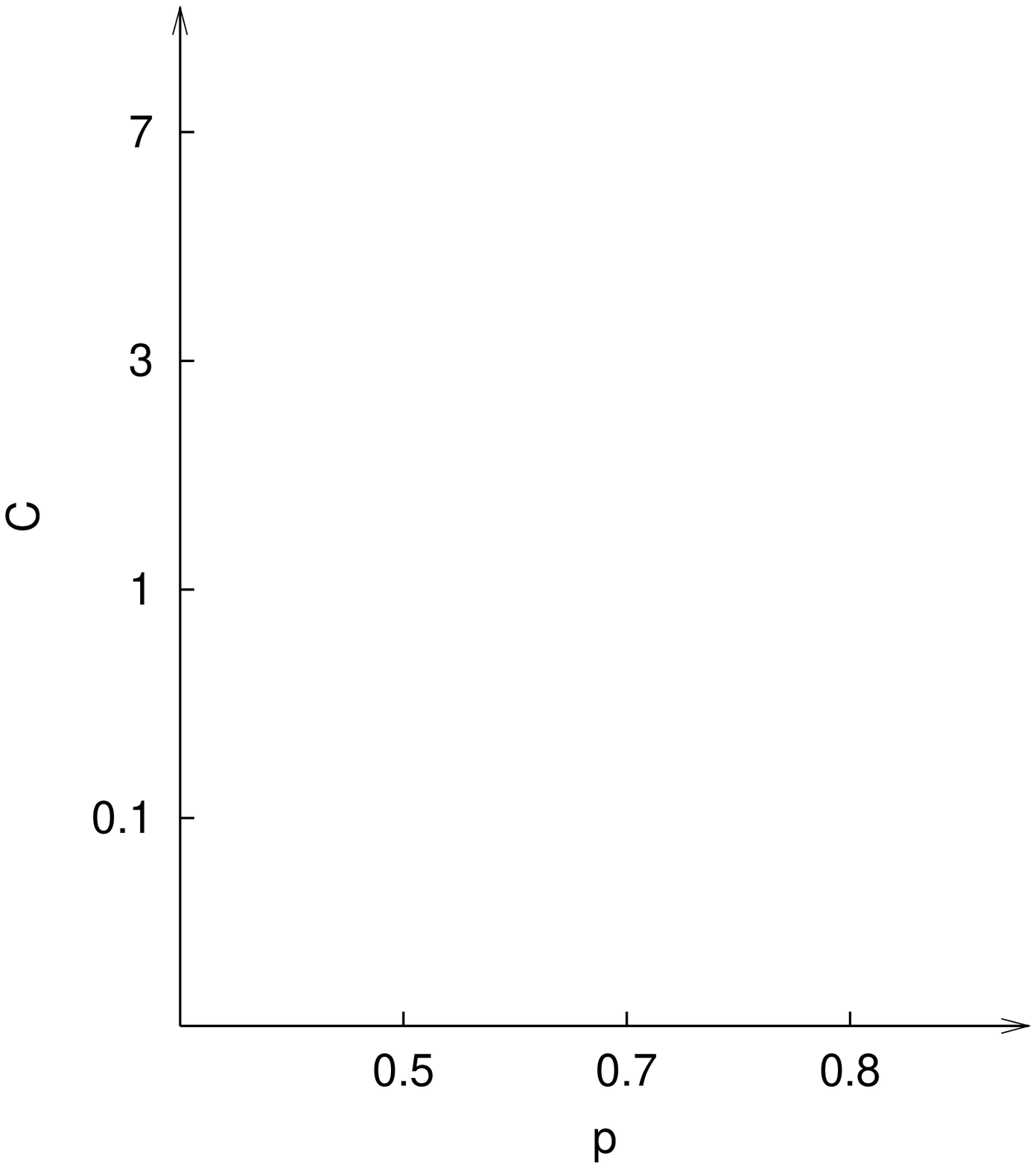}
    }

    \put(1.4,1.0){
      \includegraphics*[width=3.8in, angle=0]{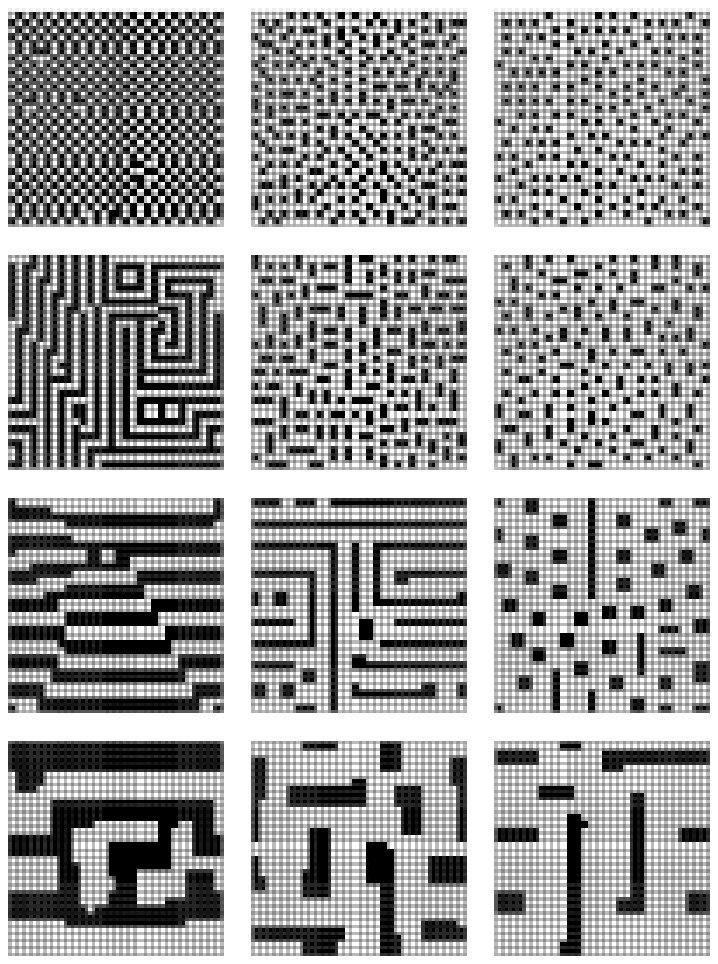}
    }
  \end{picture}
  \caption{\label{fig:pd_screened_rc3}
    Same as Fig.\ 1, but for a screened Coulomb repulsion with $r_c
    = 3a$.}
    \end{figure}

\begin{figure}
  \setlength{\unitlength}{1.0in}
  \begin{picture}(12,7)
    \put(0,0){
      \includegraphics*[width=5.8in, angle=0]{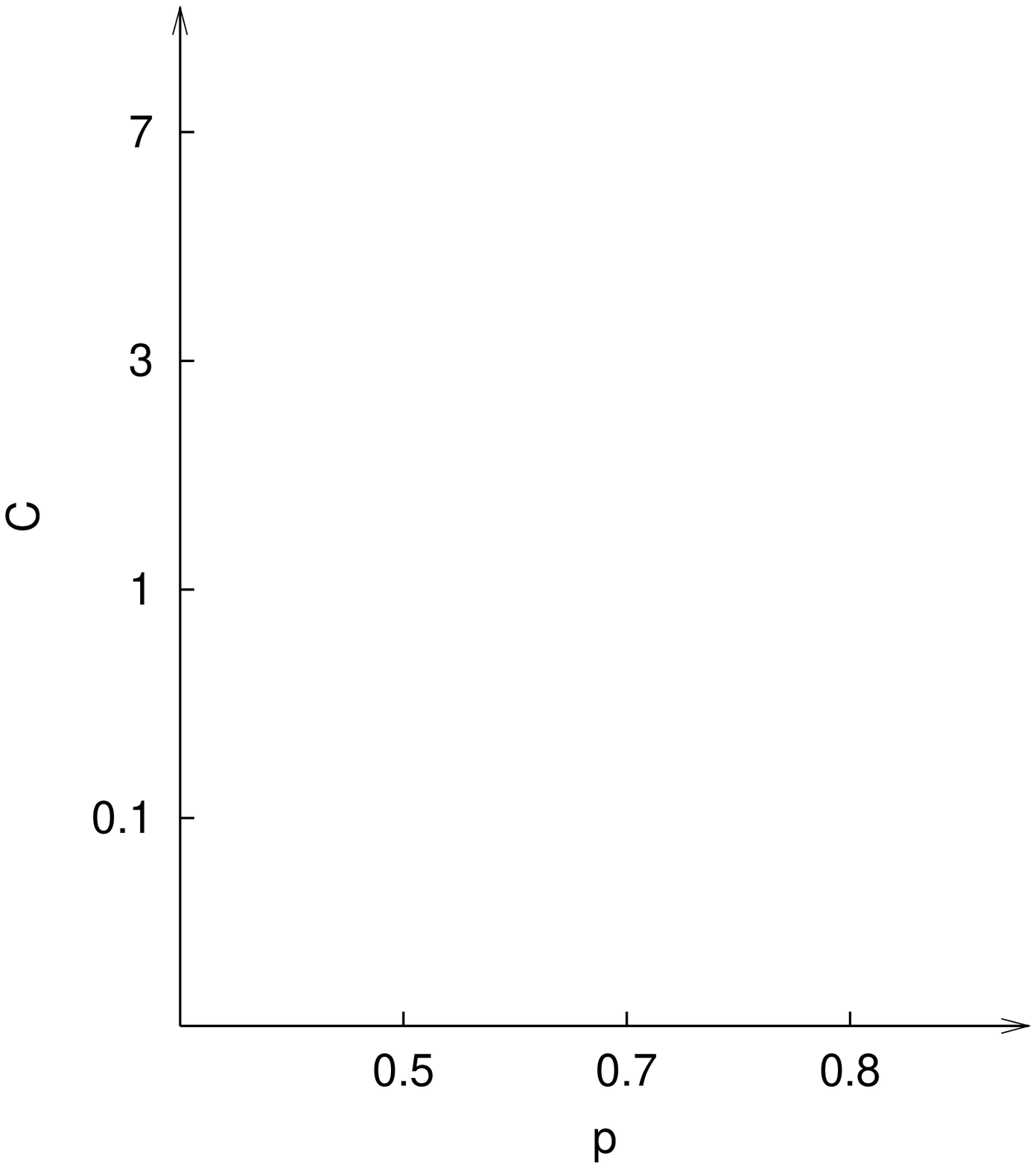}
    }

    \put(1.4,1.0){
      \includegraphics*[width=3.8in, angle=0]{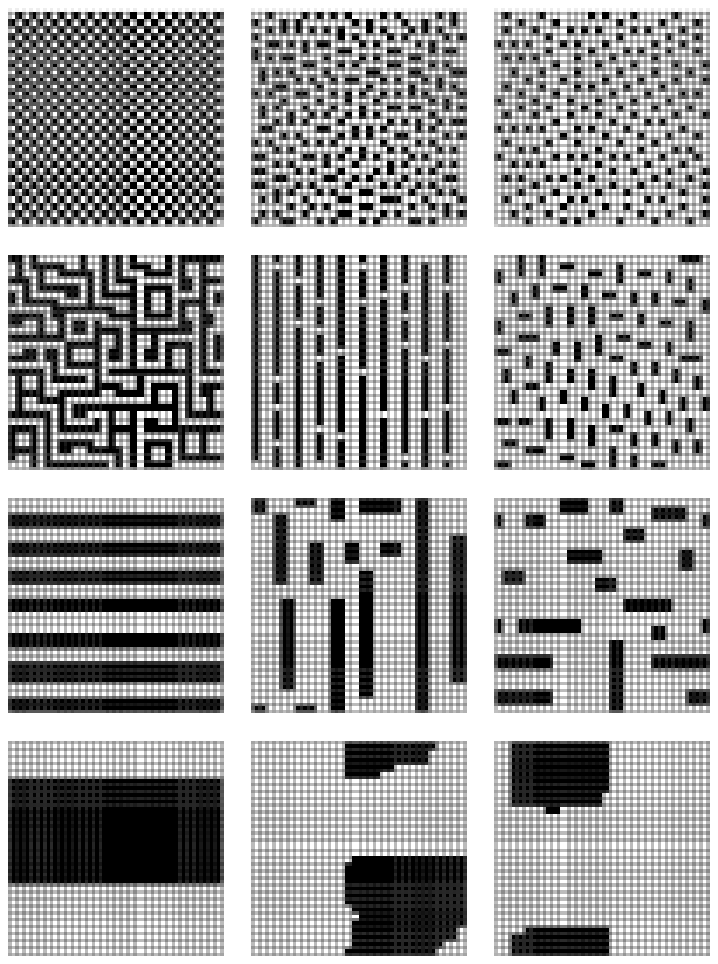}
    }
  \end{picture}
  \caption{\label{fig:pd_screened_rc1}
  Same as Fig.\ 3 but with $r_c = a$.}
    \end{figure}

%----------- COULOMB SCREENED r_c = 1, r_c = 3
Figs.~\ref{fig:pd_screened_rc3} and \ref{fig:pd_screened_rc1} show
spin configurations analogous to those of Figs.~
\ref{fig:pd_unscreened} and \ref{fig:pd_screened_rc7}. Once again,
there is a characteristic sequence of changes with increasing
Coulomb repulsion, from phase separation, to an elongated blob
phase, to a striped phase, and finally to checkerboard-like phase.
At suitable intermediate values of $C$, certain patterns are
strikingly labyrinthine, as seen, for example,
%in Figs.\
%\ref{fig:pd_screened_rc7}, \ref{fig:pd_screened_rc3}, and
%\ref{fig:pd_screened_rc1} for $p=0.7$ and $C=3$.
in Fig.~\ref{fig:pd_screened_rc7} for $p=0.7$ and $C=3$, and in
Figs.~\ref{fig:pd_screened_rc3}, and~\ref{fig:pd_screened_rc1} for
$p=0.5$ and $C=3$.

The fact that, in the stripe regimes, the stripe widths decrease
with increasing $C$ can be crudely understood from the competition
between long-range Coulomb repulsion and short-range attraction. The
latter is characterized by an energy scale $J$, while the former has
an energy scale $C/r$ for spins separated by a distance $r$.  The
two are of comparable strength for separations $r \sim C/J$.   This
equality occurs for larger $r$ as $C$ decreases - hence, to balance
the two energies, one might expect an increasing width as $C$
becomes smaller.
%DANIEL2: I added the above paragraph.  What do you think?

%---------- ANALOGY WITH CUPRATES, UNDERDOPED INSULATING AND OVERDOPED
In Table~\ref{table1}, we show the diagonal components $\gamma_{xx}$
and $\gamma_{yy}$ of the helicity modulus tensor $\gamma$ for the
low-$T$ spin configurations of Fig.~(\ref{fig:pd_screened_rc3}), as
obtained using  Eq.~(\ref{eq:conductivity}) introduced above and
discussed further below. As may be seen, the checkerboard pattern
due to strong Coulomb repulsion dramatically suppresses $\gamma(T)$
at $p = 0.5$. Since $\gamma(T)$ is proportional to the superfluid
density $n_{s}$~\cite{roddick_stroud}, these results show that a
checkerboard pattern, as expected intuitively, strongly suppress the
superfluid density.  On the other hand, a similarly strong Coulomb
repulsion at {\em higher} spin concentrations ($p = 0.7$ and $0.8$)
suppresses $\gamma(T)$ only weakly, if at all.  This behavior
loosely resembles what is seen in doped antiferromagnets, such as
the cuprate superconductors, as a function of doping. In this
analogy, we can associate the low spin concentration in our model
with the underdoped regime of the doped antiferromagnet, where it is
believed~\cite{carlson_kivelson} that strong Coulomb repulsion leads
to an insulating behavior. Likewise, we can associate with the high
spin concentration of our model the overdoped regime of the doped
antiferromagnets, where the Coulomb repulsion becomes less important
and in which they show more metallic
behavior~\cite{carlson_kivelson}.  (Of course, there is no
antiferromagnetism in our model, though there is charge ordering.)

%%%%%%%%%%%%%%%%%%%%%%%%%%%%%%%%%%%%%%%%%%%%%%%%%%%%%%%%%%%%%%%%%%%%%%%%%%%%%%%%%
\begin{table}[hbtp]
   \begin{center}
     \begin{tabular}{||c|c|c|c||}\hline\hline
%\backslashbox[48mm]{C}{p} &  0.5            &  0.7               &   0.8                  \\ \hline
%       C \ p   &  0.5            &  0.7               &   0.8                  \\ \hline
       C    &  p = 0.5           &  p  = 0.7          &   p = 0.8                  \\ \hline
       7    & (0.00,  0.00 )     & (0.30,0.30 )       &   (0.52,0.51 )         \\ \hline
       3    & (0.00,  0.07)      & (0.37,0.39 )       &   (0.49,0.54 )         \\ \hline
       1    & (0.32,  0.00 )     & (0.18,0.00 )       &   (0.42,0.59 )         \\ \hline
       0.1  & (0.26,  0.00 )     & (0.29,0.32 )       &   (0.37,0.42 )         \\ \hline\hline

     \end{tabular}
     \caption{Helicity moduli ($\gamma_{xx}$,$\gamma_{yy}$)
       for a system described by Hamiltonian~(\ref{eq:hamil1}), using
       $r_c = 3a$,
%DANIEL, add a reference to which Hamiltonian here.
%DR. STROUD, done.
       at $T = 0.025$. $C$ is the strength of the repulsion, and $p$
       is the spin concentration.  Some snapshots of the
       system are shown in
       Fig.\ \ref{fig:pd_screened_rc3}.}\label{table1}
   \end{center}
 \end{table}

%------------- COMPARISON TO OTHER MODELS AND TO EXPERIMENTS
All of the patterns of Figs.
\ref{fig:pd_unscreened}-\ref{fig:pd_screened_rc1} (stripes,
checkerboards, and labyrinths) have been observed, both in
experiments and in simulations, in systems with competing long and
short range interactions.  For example, labyrinthine structures have
been observed in experiments on magnetic garnets~\cite{albuquerque},
%DANIEL, are these experiments?
%DR. STROUD, yes, I added the word 'experiments'
and they have been obtained in simulations of spin-1 Ising
%DANIEL, I think these Hamiltonians are called something other than
%Ising-like.
%DR. STROUD, Iglesias calls his model Hamiltonian a ``spin-1 Ising-like Hamiltonian''
% should we call it ``spin-1 Ising Hamiltonian''? or something else?
Hamiltonians with competing long-range and short-range
interactions~\cite{iglesias}.
%DANIEL, do these have competing long and short range interactions.
%DR. STROUD, yes, I added this detail above
Striped magnetic phases have been observed experimentally in
ferrimagnetic garnet films~\cite{seul_wolfe}, and obtained in
simulations of spin-1/2 Ising models with a long-ranged dipolar
interaction~\cite{booth}. Checkerboard patterns in the low-$T$
electronic structure of Ca$_{2-x}$Na$_{x}$CuO$_{2}$Cl$_{2}$ have
been experimentally observed~\cite{hanaguri_davis}, and also
obtained numerically in simulations of a classical spin-1 lattice
gas model with short-range ferromagnetic coupling and long-range
antiferromagnetic Coulomb interactions~\cite{low_emery}.   There
have also been numerical
%DANIEL, I added the word "numerical" here - is this correct?
%DR. STROUD, yes, it's correct
studies~\cite{stoycheva} of the stripe melting transition in systems
governed by Ising-1/2 models with short ranged ferromagnetic and
long ranged antiferromagnetic couplings.   The present model differs
from all of these in having an XY, rather than an Ising, attractive
interaction between the spins.  In order to unambiguously
distinguish our model from all these others, it would be desirable
to carry out simulations on our model I but with an {\em
un-truncated} Coulomb repulsion.

%%%%%%%%%%%%%%%%%%%%%%%%%%%%%%%%%%%%%%%%%%%%%%%%%%%%%%%%%%%%%%%%%%%%%%%%%

\subsection{Model II}

\subsubsection{Numerical Results}

In our studies of Model II, we have emphasized the $T$-dependence of
various quantities, in particular $\gamma(T)$, as well as
$S_{1}(\vec q, T)$, and $S_{2}(\vec q, T)$ for special values of
$\vec{q}$.  We will also present snapshots of spin configurations at
various $T$'s.

\begin{figure}\centering

  {\includegraphics[width=7cm]{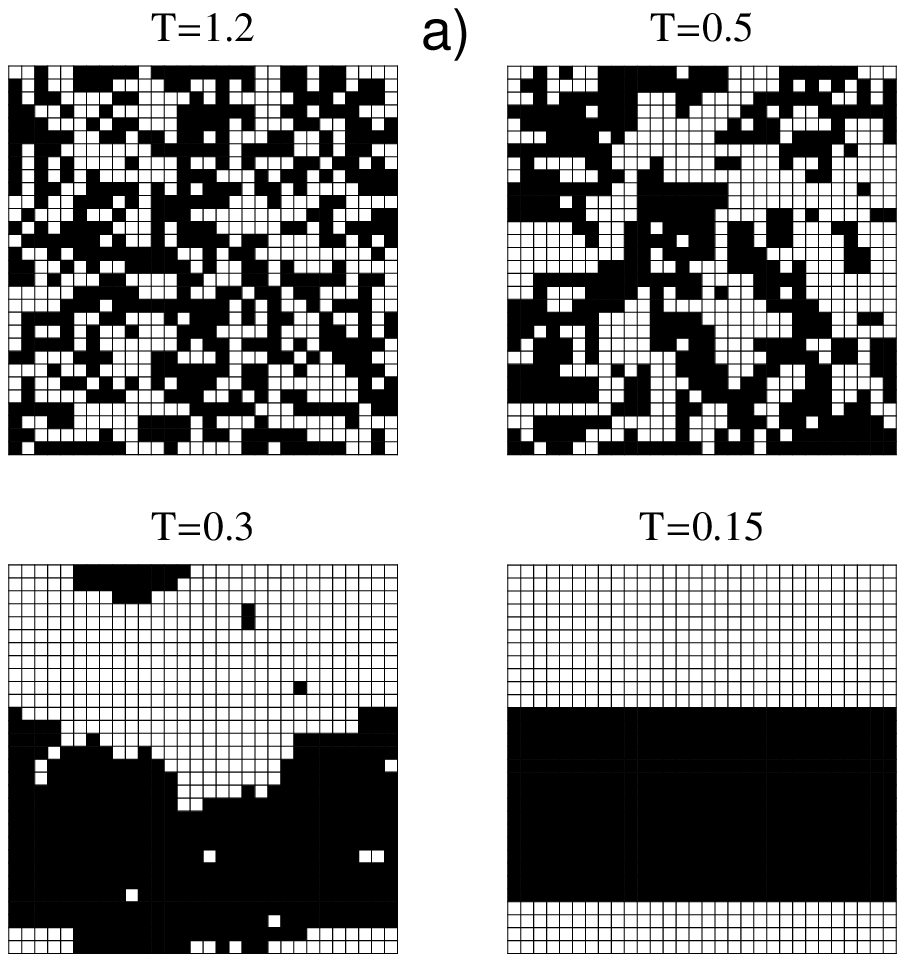}}
  \qquad\qquad\quad
  {\includegraphics[width=8cm]{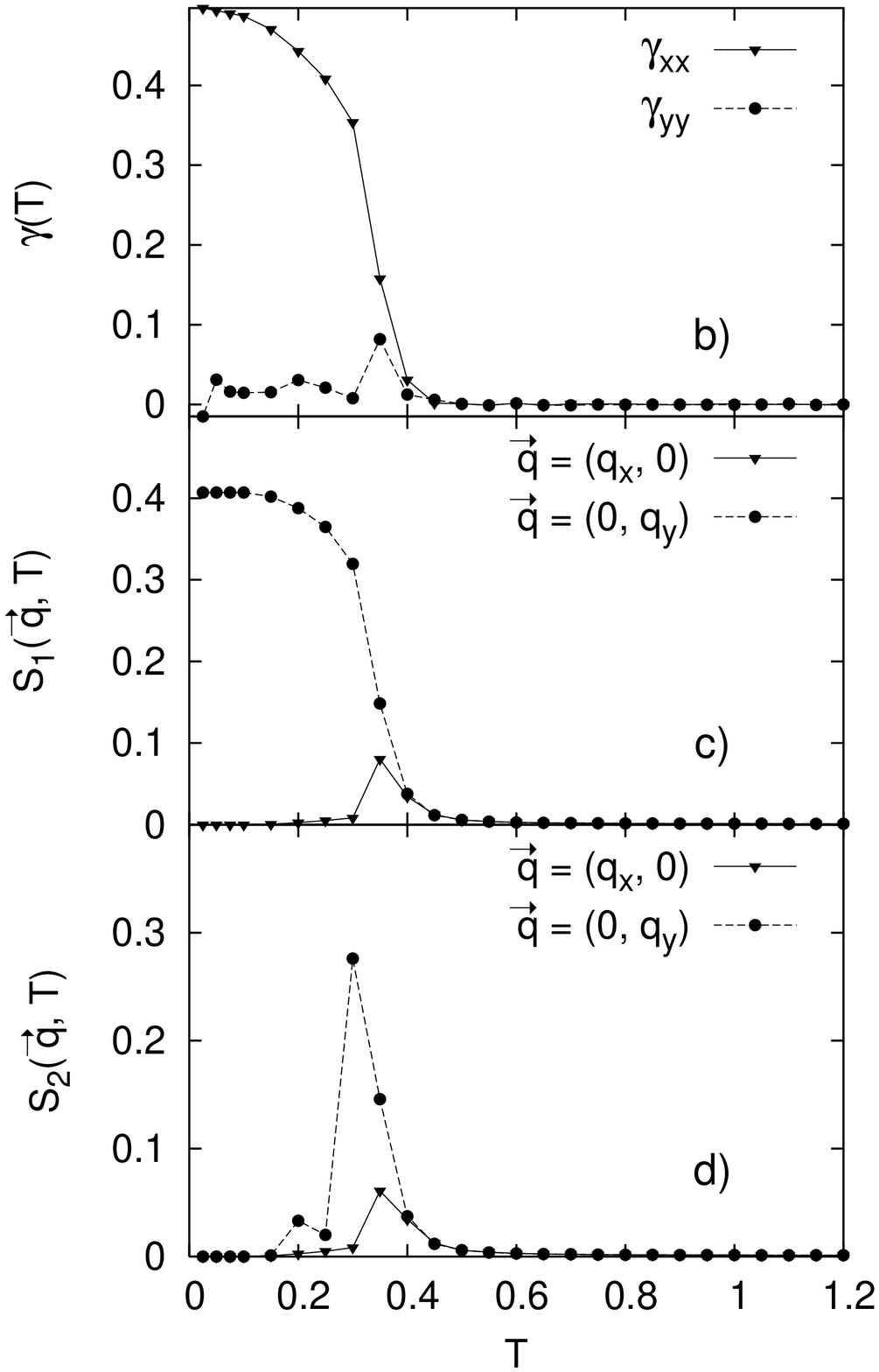}}
  \caption{Some results for Model II, with $p = 0.5$, and weak second nearest
    neighbor repulsion $B = 0.1$.  (a)~Representative snapshots of
    spin configurations at different temperatures T. White (black)
    squares are occupied (vacant) sites.  (b)~Diagonal components
    $\gamma_{xx}$ and $\gamma_{yy}$ of the helicity modulus tensor,
    exhibiting a coherence transition. (c)~and (d):~Order parameters
    sensitive to phase separation, with $q_{x} = 2 \pi / N_{x} a$ and
    $q_{y} = 2 \pi / N_{y} a$.  }
  \label{fig:p0.5_nn2_B0.1}
\end{figure}

% -------------------  WEAK NN p = 0.5
Fig.~\ref{fig:p0.5_nn2_B0.1}a shows of snapshots of the spin
configurations for $p = 0.5$ at different $T$ for Model II with $A =
0$,
%DANIEL, do you mean $A = 0$?
% DR. STROUD, yes, I corrected
 $B = 0.1$ (weak second-nearest-neighbor repulsion). At $T = 1.2$, the
 spins are already tending to form clusters of
various sizes and shapes, a tendency which is clearer at $T = 0.5$.
At this latter $T$, even though the spins form a connected path in
the horizontal ($x$) direction, $\gamma_{xx}(T)$ (shown in
Fig.\ref{fig:p0.5_nn2_B0.1}b) remains small, indicating no phase
coherence.  By $T = 0.3$, the system has phase separated into two
large domains, made up of spins and vacancies, respectively. This
transition to a phase separated configuration is signaled by
non-zero values of $S_{1}(\vec q, T)$
(Fig.~\ref{fig:p0.5_nn2_B0.1}c) for $T < 0.3$ and by a peak in
$S_{2}(\vec q, T)$ (Fig.~\ref{fig:p0.5_nn2_B0.1}d) near $T = 0.3$,
with $\vec q= 2 \pi \hat y / N_{y} a$.   The transition to the
phase-coherent state, signaled by the finite value of $\gamma(T)$,
and that to the phase separated state, seem to occur at the same
$T$.

These results show that the annealed system has strikingly different
behavior from a system with quenched disorder.  In the latter case, $p
= 0.5$ is below the critical percolation threshold $p_c \simeq 0.59$ for 2D
site diluted square lattice~\cite{sahimi}.
%DANIEL, need a reference here.
%DR. STROUD, I added the reference
Hence, there is no phase coherence at any $T$ in the quenched case.
However, for annealed disorder, the phase separation, which occurs
when the Coulomb repulsion is weak enough, leads to an infinite
percolating cluster of spins, and hence to phase coherence in one of
the two principal directions, even at a spin concentration below
$p_c$.  Although similar phase separation occurs in Model I, it does
not appear to lead to an infinite cluster of spins for $p < 0.5$, and
hence, there is no $T = 0$ phase coherence in that case.

\begin{figure}\centering

  {\includegraphics[width=7cm]{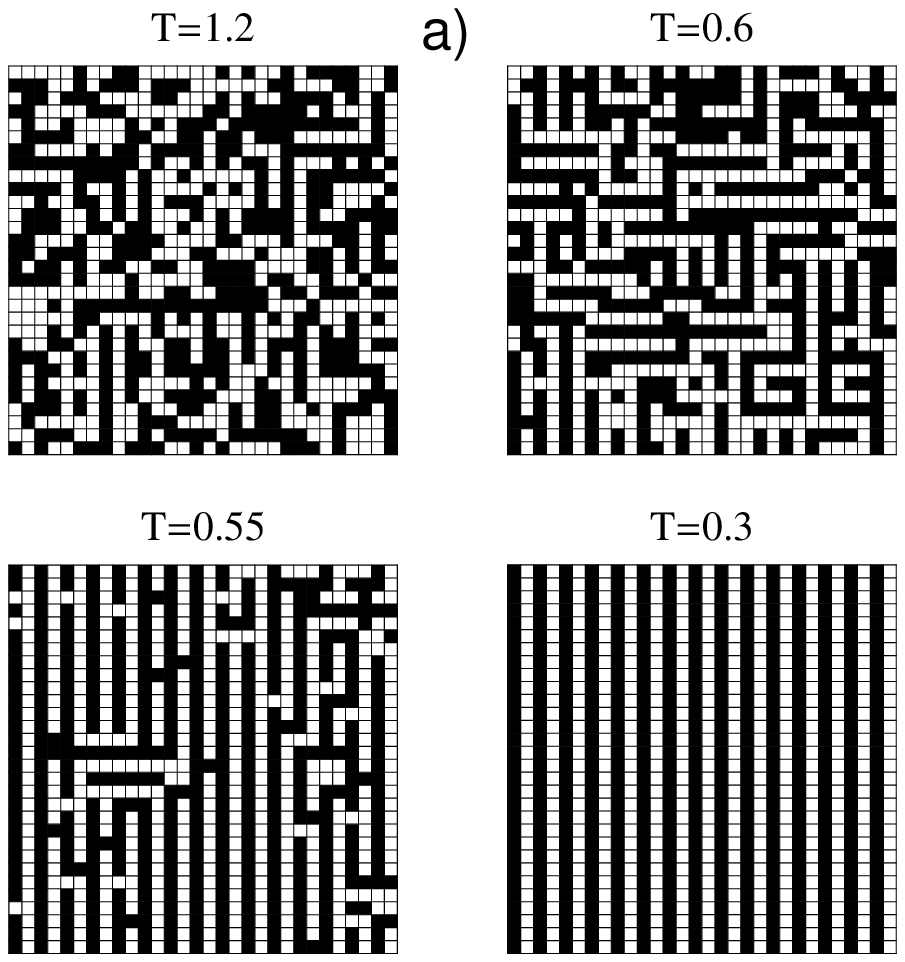}}
  \qquad\qquad\quad
  {\includegraphics[width=8cm]{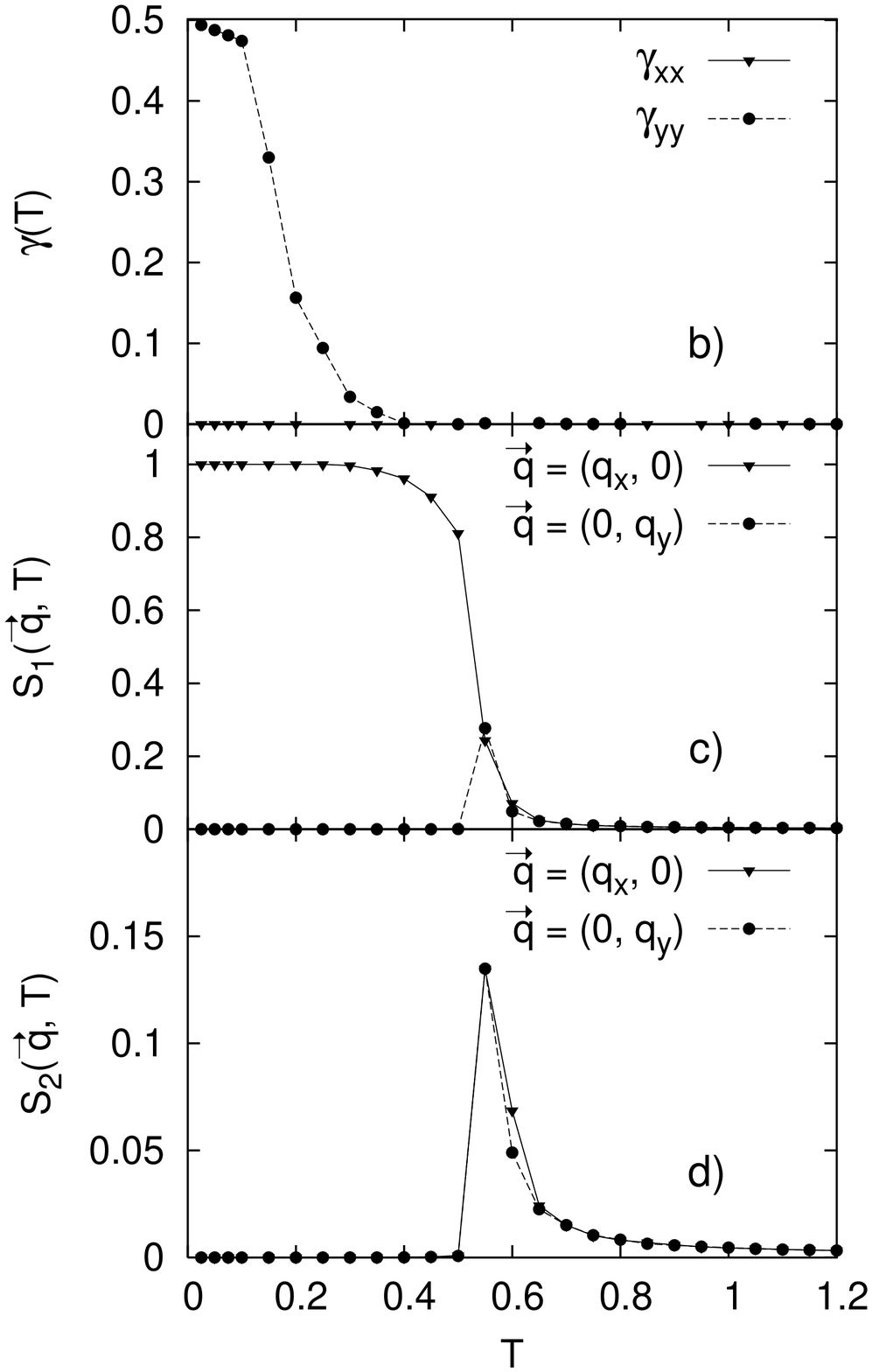}}
  \caption{
    Results for Model II, with $p = 0.5$, and strong second nearest
    neighbor repulsion ($B = 1.0$).  (a)~Representative snapshots of spin configurations at different
    temperatures T. White (black)
    squares are occupied (vacant) sites. (b)~Diagonal components $\gamma_{xx}$ and
    $\gamma_{yy}$ of the helicity modulus tensor.
    %DANIEL,  maybe you want to change the legend so it shows gamma_{xx} instead of \gamma_x, etc.
    %DR. STROUD, done.
    transition. (c)~and (d):~Order parameters sensitive to stripe formation, with
    $q_{x}=\pi / a$, $q_{y}=\pi / a$.  }
  \label{fig:p0.5_nn2_B1.0}
\end{figure}

% ------------------- STRONG NNN p = 0.5
Fig.~\ref{fig:p0.5_nn2_B1.0} shows $p = 0.5$ for the same model but
with a stronger second-neighbor repulsion ($A = 0$; $B = 1.0$).
Fig.~\ref{fig:p0.5_nn2_B1.0}a  shows that at $T = 1.2$ and
especially at $T = 0.6$, there is a tendency for stripe formation,
though the stripes remain of finite length.  For $T = 0.55$ the
system transforms to a mostly unidirectional striped phase. This
transition is clearly signaled by a finite value in $S_{1}(\vec q, T)$
for $T < 0.55$ (Fig.~\ref{fig:p0.5_nn2_B1.0}c), and by a peak in
$S_{2}(\vec q, T)$ (Fig.~\ref{fig:p0.5_nn2_B1.0}d) near $T = 0.55$, for
$\vec q=\pi \hat x / a$.   At $T = 0.3$, the system forms an ordered
array of stripes in the vertical ($y$) direction, manifested in the
fact that $S_{1}(\vec q, T) = 1$ at and below this $T$.  In contrast to
the case of phase separation obtained for weak repulsion in
Fig.~(\ref{fig:p0.5_nn2_B0.1}), for which the phase coherence has a
sharp onset temperature which coincides with the transition to a
phase separated state, the phase-coherent state in the present case
has a more gradual onset with decreasing $T$, and occurs at a lower
$T$ than the transition to a striped phase.  Also, the coherence
transition is observable in only one of the two principal
directions, as expected since phase coherence is geometrically
impossible in the direction perpendicular to the stripes.

% ------------------- STRONG NN A= 1.5

\begin{figure}\centering
  {\includegraphics[width=7cm]{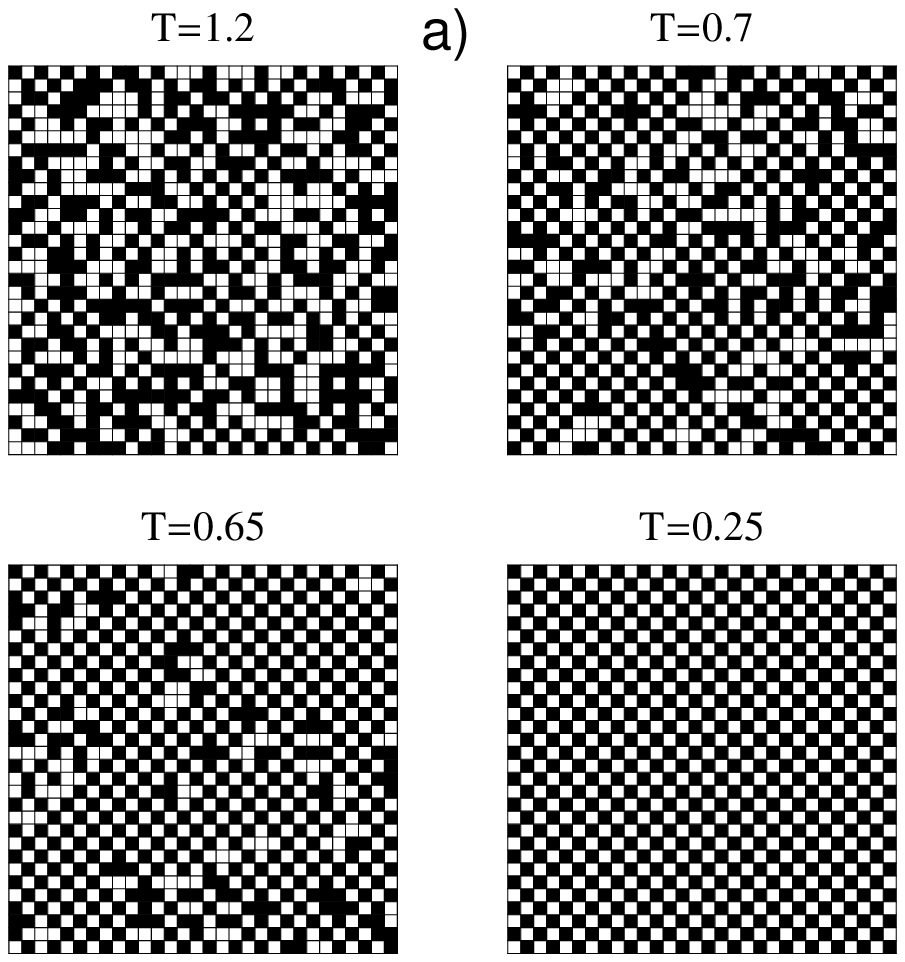}}
  \qquad\qquad\quad
  {\includegraphics[width=8cm]{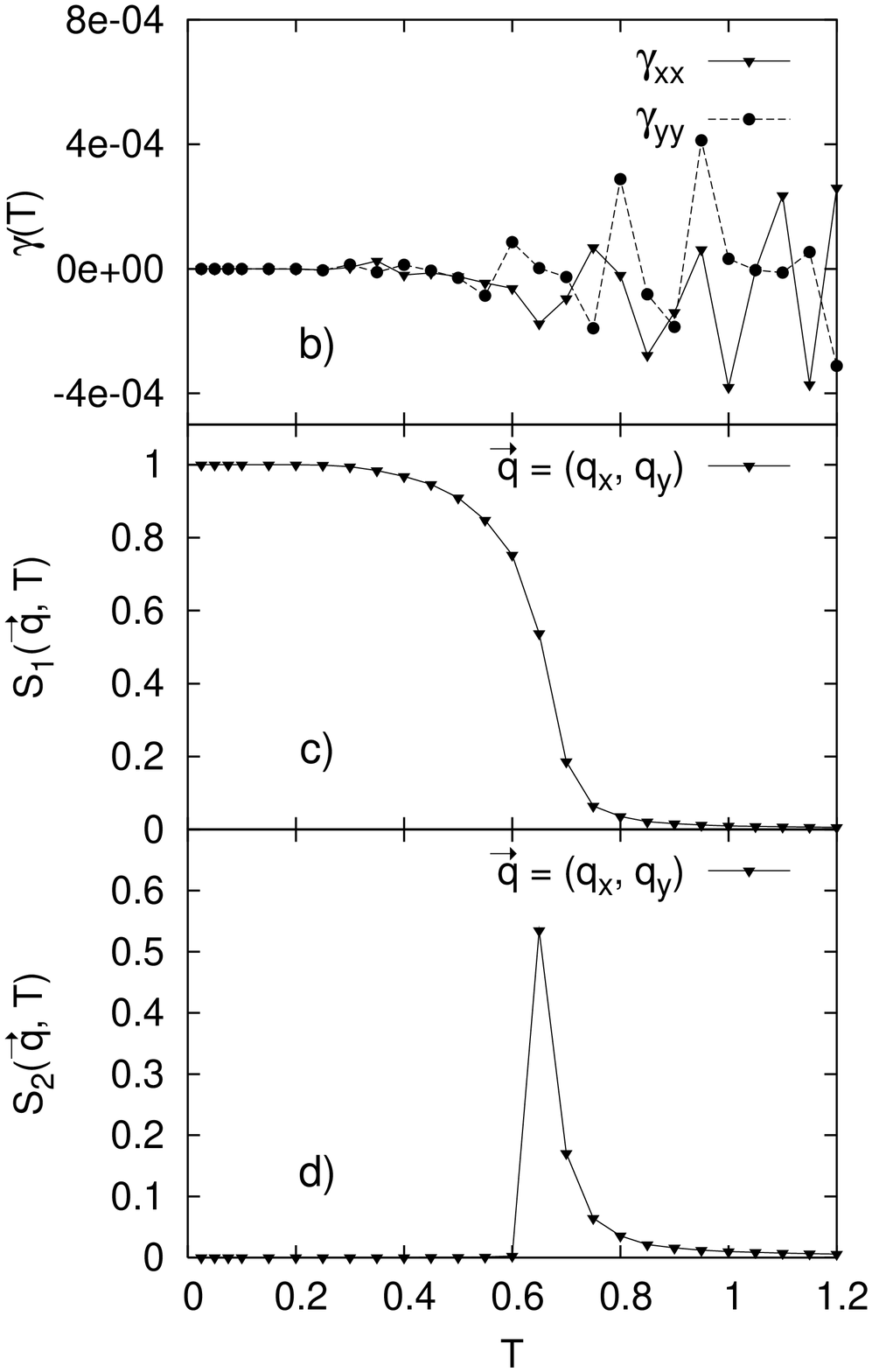}}
  \caption{
    Results for Model II, with $p = 0.5$, and a strong
    nearest-neighbor repulsion ($A = 1.5$).  (a)~Representative snapshots of spin
    configurations at different
    $T$.  White (black) squares are
    occupied (vacant) sites. (b)~Diagonal components $\gamma_{xx}$ and $\gamma_{yy}$ of helicity modulus tensor.
    (c)~and (d)~Order parameters sensitive to checkerboard order, with
%DANIEL, you originally wrote phase separation, but I changed it,
    $\vec q=\pi (\hat x +\hat y) / a$.   }
\label{fig:p0.5_nn3_A1.5}
\end{figure}

We turn now to results for Model II with nearest-neighbor repulsion
only.  Fig.~\ref{fig:p0.5_nn3_A1.5} shows results for $A = 1.5$, $B
= 0.0$, and $p = 0.5$. As in the cases described above, the system
goes from a disordered phase at high $T$, to an ordered low-$T$
phase.  However, Fig.~\ref{fig:p0.5_nn3_A1.5}b shows that the system
is prevented at all $T$ from undergoing a transition with a non-zero
$\gamma$.  This can be understood on simple geometrical grounds.
Fig.~\ref{fig:p0.5_nn3_A1.5}a shows that the strong nearest-neighbor
repulsion decreases the number of nearest-neighbor spins at low $T$.
Since such nearest-neighbors are essential in for an attractive
interactions between the nearest-neighbor XY spins, this reduction
insures that $\gamma(T) \sim 0$, as we observe numerically. The
transition to a checkerboard-like spin pattern is indicated by the
finite value of $S_{1}(\vec{q}, T)$ for $T < 0.65$ and the peak in
$S_{2}(\vec{q}, T)$ near $T = 0.65$, for $\vec{q} = (\pi/a, \pi/a)$.
%DANIEL, is this the right q value here?
% DR. STROUD, I corrected by the factor 1/a

\begin{figure}\centering
  {\includegraphics[width=7cm]{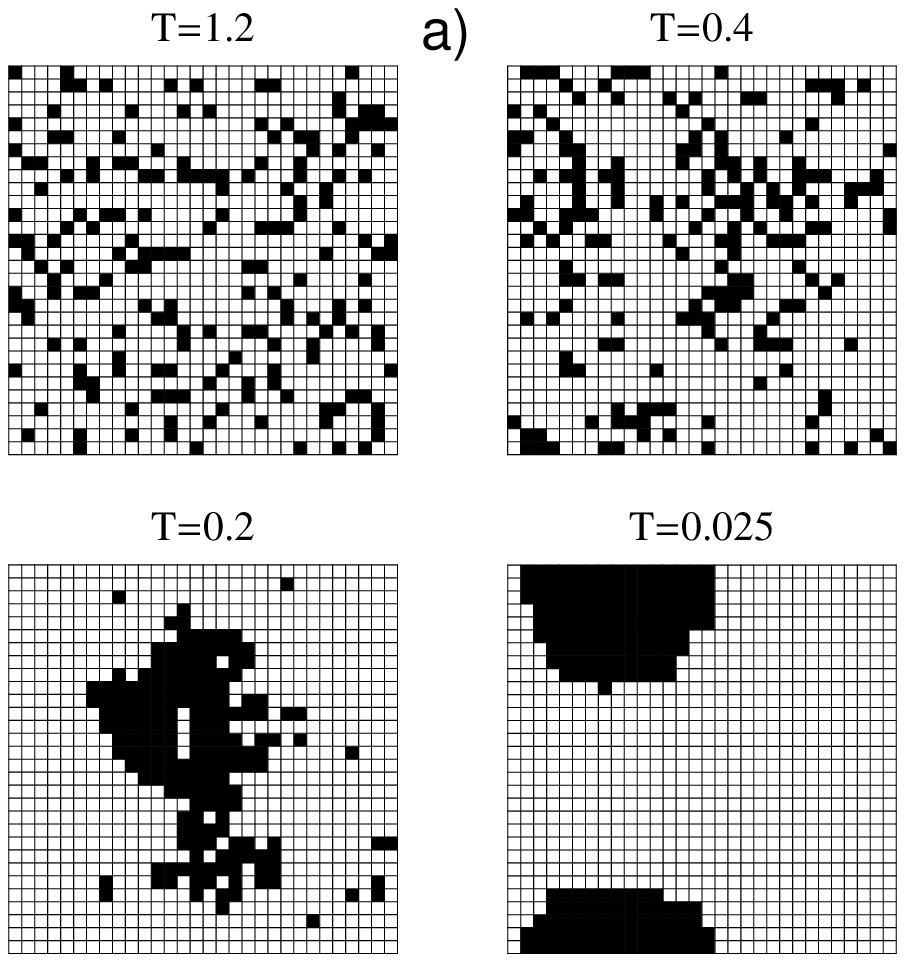}}
  \qquad\qquad\quad
  {\includegraphics[width=8cm]{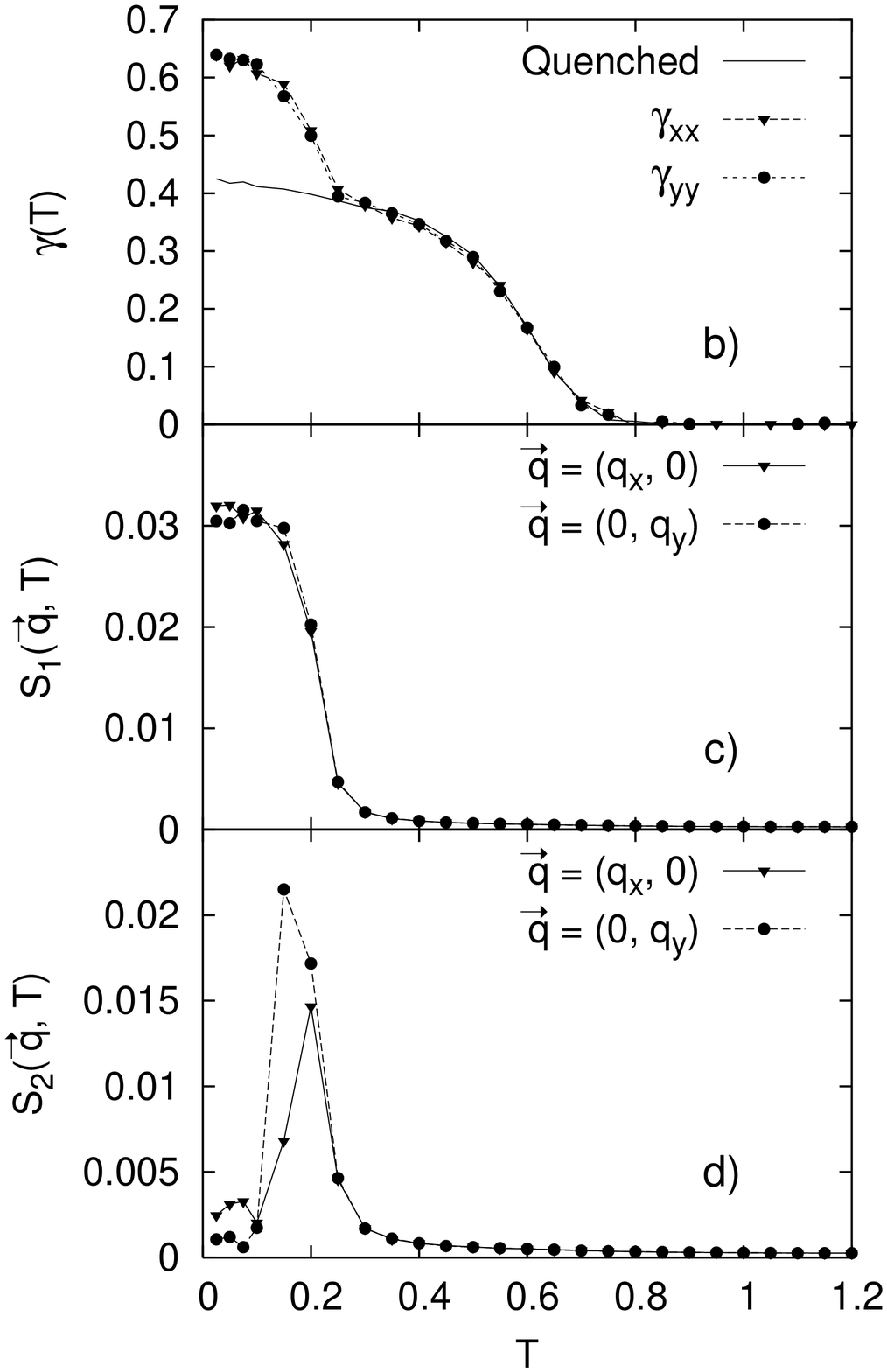}}
  \caption{Results for Model II, with $p = 0.8$, and weak nearest
    neighbor repulsion ($A = 0.5$).  (a)~Spin snapshots at various
    temperatures $T$. White (black) squares are occupied (vacant) sites. (b)~Diagonal components of helicity modulus tensor.
    (c)~and (d)~Order parameters sensitive to phase separation, with
    $q_{x} = 2 \pi / N_{x} a$ and $q_{y} = 2 \pi / N_{y}
    a$. }
  \label{fig:p0.8_nn3_A0.5}
\end{figure}

% ======================   p = 0.8
We now present results for Model II with $p = 0.8$.  Since $p$
exceeds the site percolation threshold of $p_{c} \simeq 0.59$,
$\gamma(T)$ should become nonzero at sufficiently low $T$.
% ------------------- WEAK NN  p = 0.8
Fig.~\ref{fig:p0.8_nn3_A0.5} shows results for weak nearest-neighbor
repulsion ($A = 0.5$; $B = 0.0$), and $p = 0.8$. Fig.\
\ref{fig:p0.8_nn3_A0.5}(b) shows that $\gamma_{xx}(T) \sim
\gamma_{yy}(T) $ for all $T$. We denote the helicity modulus in this
isotropic regime simply as $\gamma(T)$. $\gamma(T)$ increases
smoothly with decreasing $T$ to $T \sim 0.25$, and, in this
regime, coincides with $\gamma(T)$ for quenched disorder at same
spin concentration (continuous line with no symbols in
Fig.~\ref{fig:p0.8_nn3_A0.5}b).  However, the quenched and
annealed results differ for $T < 0.25$.  $\gamma(T)$ for the
annealed case changes slope near $T = 0.25$ because the spins
phase-separate.   This phase separation is clearly visible in
Fig.~\ref{fig:p0.8_nn3_A0.5}a, which shows snapshots of the spin
configurations for several values of $T$.  It is also signaled by the finite
value of $S_{1}(\vec{q}, T)$ for $T \leq 0.2$, and a peak in
$S_2(\vec{q}, T)$  around $T = 0.2$
(Figs.~\ref{fig:p0.8_nn3_A0.5}c and~\ref{fig:p0.8_nn3_A0.5}d),
for $\vec q= 2 \pi \hat x / (N_{x} a)$, and $\vec q= 2 \pi \hat y /
(N_{y} a)$.

\begin{figure}\centering
  {\includegraphics[width=7cm]{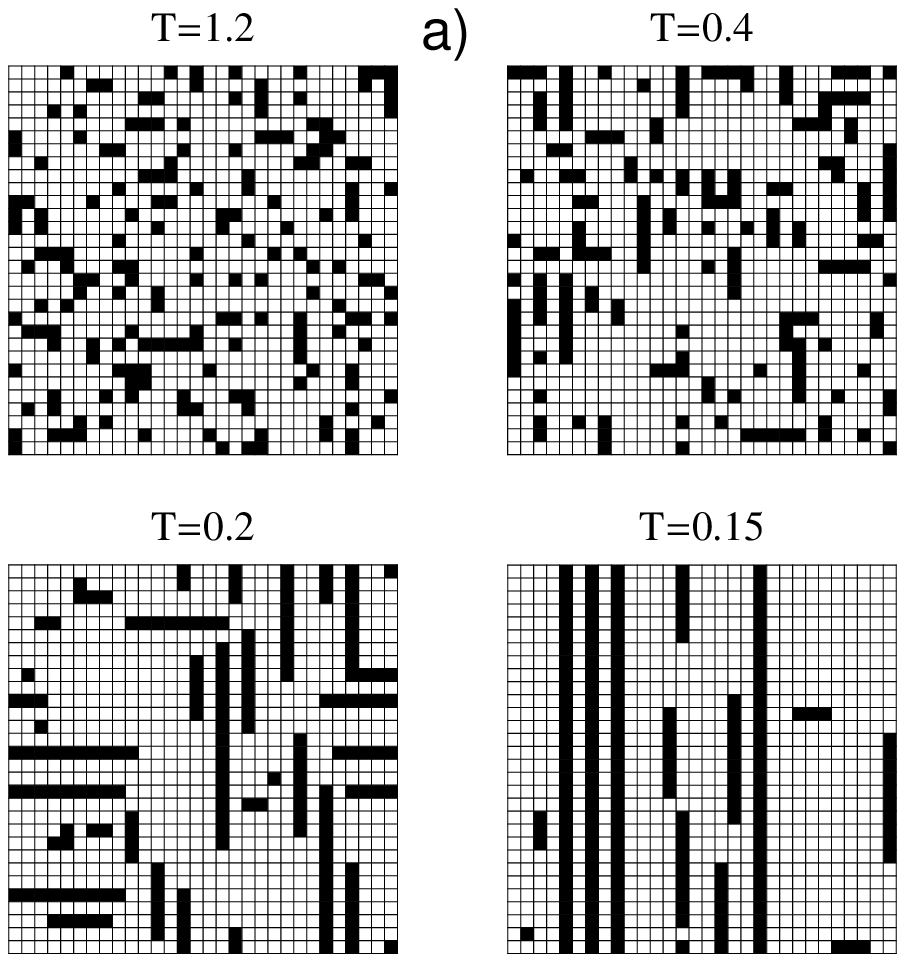}}
  \qquad\qquad\quad
  {\includegraphics[width=8cm]{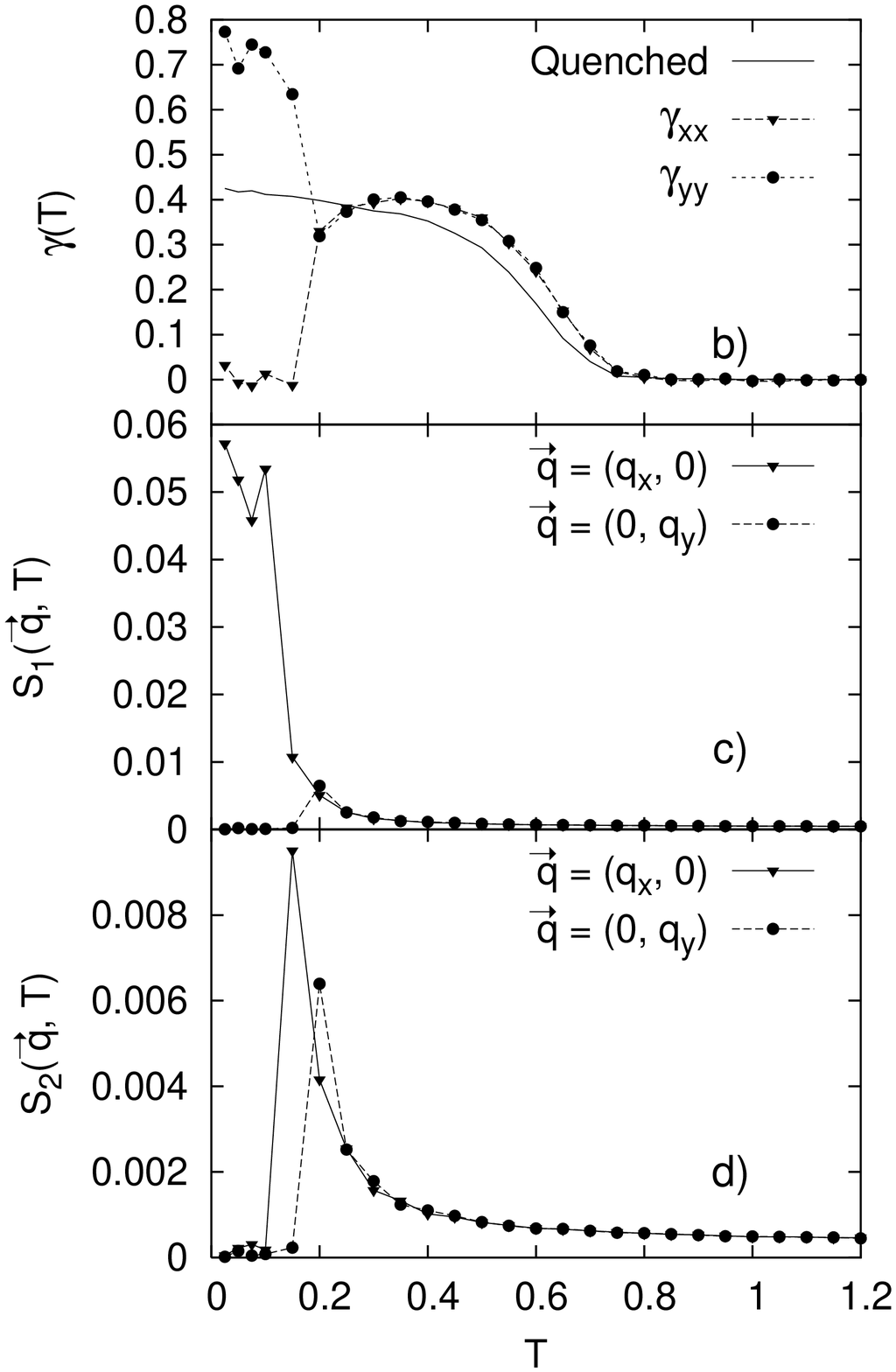}}
  \caption{Results for Model II, with $p = 0.8$, and strong second nearest
    neighbor repulsion ($B = 1.0$).  (a)~Spin snapshots at different
    temperatures $T$. White (black)
    squares are occupied (vacant) sites. (b)~Diagonal components of helicity modulus tensor. (c)~and (d):~Order parameters
    sensitive to stripe formation, with
    $q_{x}=\pi / a$, $q_{y}=\pi  / a$.  }
  \label{fig:p0.8_nn2_A1.0}
\end{figure}

% ------------------- STRONG NNN  p = 0.8
Fig.~\ref{fig:p0.8_nn2_A1.0} shows the analogous results for
stronger nearest-neighbor repulsion (A = 0, B = 1, and p = 0.8). In
this case, the vacancies tend to cluster into {\em stripes}, not
blobs - which become longer as $T$ is decreased from $1.2$ to $0.4$;
below about $T = 0.2$, the stripe pattern becomes anisotropic.  The
helicity moduli $\gamma_{xx}(T)$ and $\gamma_{yy}(T)$ become nonzero
around $T = 0.7$ but remain {\em isotropic} (i.\ e., nearly equal)
down to around $T = 0.2$.  For $T < 0.2$, the anisotropic stripe
pattern leads to a dramatic anisotropy in the helicity moduli:
$\gamma_{xx}(T)$ falls to 0, but $\gamma_{yy}(T)$ increases. This
behavior has an obvious geometrical explanation in the long stripes
parallel to $y$, which inhibit phase coherence in the $x$ direction.
At the lowest temperature of $T=0.025$, there are several horizontal
stripes spanning the sample, which cause $\gamma_{xx}$ to vanish,
while $\gamma_{yy}$ attains a value well above $\gamma(T)$ in the
case of quenched disorder. The transition to a striped phase is
observed in the onset of a finite value of $S_{1}(\vec{q}, T)$, as
well as in the peak in $S_{2}(\vec{q}, T)$, around $T = 0.2$, for
$\vec{q} = (\pi/a, 0)$.
%[CHECK q value]
%DANIEL, what q value here?
% DR. STROUD, I answered this above
In contrast to the case of weak nearest-neighbor repulsion shown in
Fig.~\ref{fig:p0.8_nn3_A0.5}, where $\gamma(T)$  closely resembles
that of the system with quenched disorder for all $T \geq 0.25$, the
effect of strong second nearest-neighbor repulsion here leads to a
{\em non-monotonic} $\gamma(T)$: $\gamma(T)$ is largest at
intermediate $T$, and small at high $T$, where it is destroyed by
thermal fluctuations, and at low $T$, where it is frustrated by the
formation of stripes.  By contrast, at intermediate $T$, short
stripes have formed but since they are short and randomly oriented,
they are insufficient to prevent a finite $\gamma(T)$.

\begin{figure}\centering
  {\includegraphics[width=7cm]{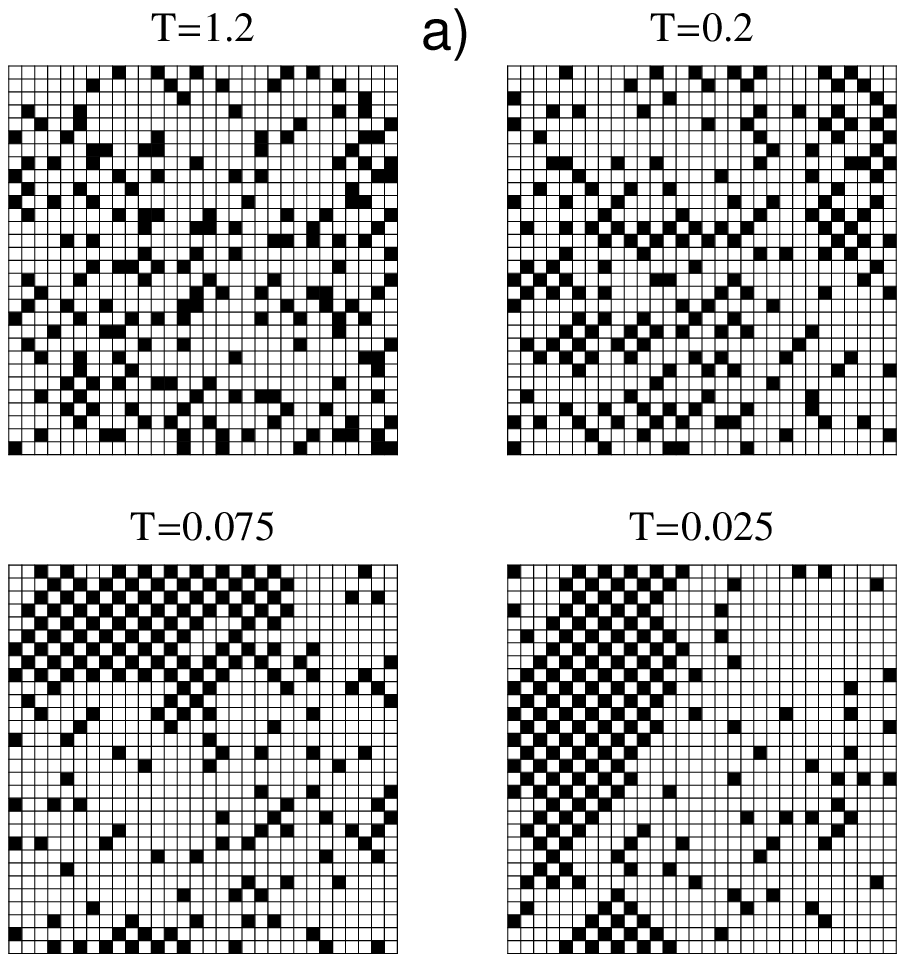}}
  \qquad\qquad\quad
  {\includegraphics[width=8cm]{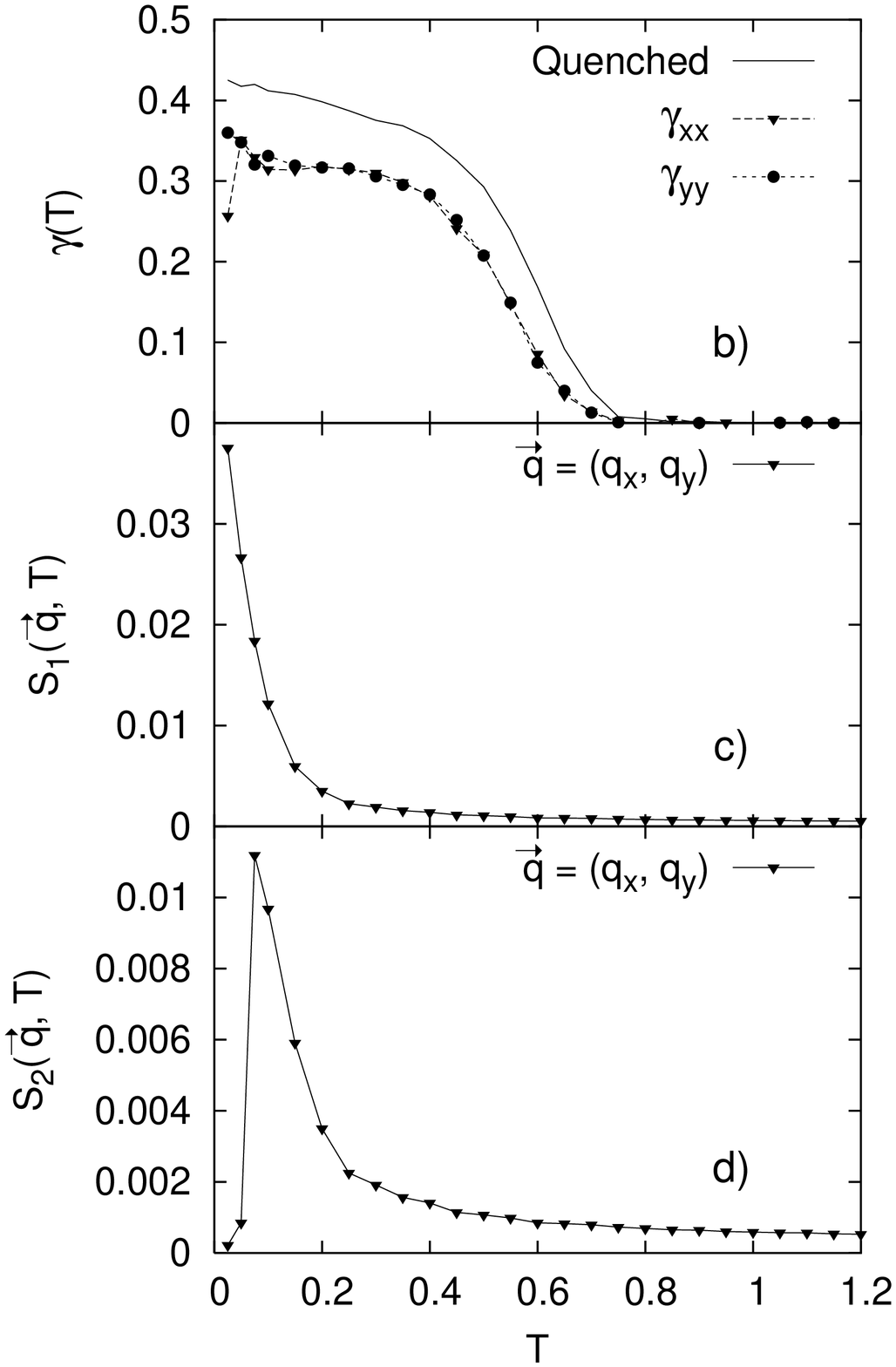}}
  \caption{Model II results, with $p = 0.8$, and a strong
    nearest-neighbor repulsion ($A = 1.5$).  (a)~Spin snapshots at different
    various temperatures $T$. White (black) squares are
    occupied (vacant) sites. (b)~Diagonal components of the helicity modulus tensor.
    (c)~and (d):~Order parameters sensitive to checkerboard order, with
    $\vec q=\pi (\hat x +\hat y) / a$.   Note the apparent phase separation at low $T$ into a checkerboard
    phase and a fully occupied spin phase.}
  \label{fig:p0.8_nn3_A1.5}
\end{figure}

%------------------- STRONG NN  p = 0.8
Finally, we show the results for strong nearest-neighbor repulsion
(A = 1.5, B = 0, p = 0.8) in Fig.\ ~\ref{fig:p0.8_nn3_A1.5}. Similar
As in the case of nearest-neighbor repulsion at $p = 0.5$ shown in
Fig.~\ref{fig:p0.5_nn3_A1.5}, where the repulsion prevented phase
ordering at all $T$,  $\gamma_{xx}(T)$ and $\gamma_{yy}(T)$ are
substantially {\em reduced} compared to the corresponding quenched
values, and for the same reason: the nearest-neighbor repulsion
tends to decrease the number of spin nearest-neighbors, which in
turn decreases the tendency of the system towards phase coherence.
The Figure also makes apparent that an "antiferromagnetic"
(checkerboard) pattern of $n_i$'s is emerging at low $T$, which
competes with the XY transition.  But at this $p$, in contrast to $p
= 0.5$, clumps of checkerboard-ordered regions of zero helicity
modulus can coexist with regions of finite helicity modulus, leading
to a nonzero global value of $\gamma(T)$.

%DANIEL, were such simulations done previously at A = 0?
% (Just plain annealed disorder)?  I just assumed it was known.
% DR. STROUD, I looked on the APS journals and didn't find any study
% of XY models with annealed disorder.

%-------------------------------------------------------------------

\subsubsection{Analytical Results}

At $T = 0$, it is possible, by a simple comparison of energies, to
calculate {\em analytically} the critical values $A_c$ and $B_c$ at
which the system changes from phase-separated to checkerboard or
striped order.   In the phase-separated state, all the spins are
contained in clusters in which all sites are occupied. Thus, the
energy per spin in this state (assuming $B = 0$) is simply
\begin{equation}
E^{ps} = -2J + 2A, \label{eq:psep}
\end{equation}
where we have used the fact that each spin has two nearest-neighbor
spins.  For the checkerboard ground state (taking $B = 0$), the
ground-state energy per spin is simply
\begin{equation}
E^{check} = 0. \label{eq:check}
\end{equation}
The critical value of $A_c$ is just that value of $A$ where the two
energies are equal, i.\ e., $A_c = J$.  For $A
> A_c$, the ground state is checkerboard; for $A < A_c$, it is phase
separated.  Our simulations agree with this analytical prediction.

%DANIEL, does this agree with your simulations?
% DR. STROUD, Yes.

%-----------------------------------------------------------------
To calculate $B_c$, we assume $A = 0$ and compare the energies of
the two spin arrangements at $T = 0$, using a simple bond-counting
argument. In the phase-separated state, at $T = 0$, the ground state
consists of large blobs of spins and vacancies.  Disregarding the
surface energies, we find that the energy {\em per spin} is
\begin{equation}
E^{ps} = -2J + 2B, \label{eq:Eps}
\end{equation}
since each spin has two nearest and two second-nearest spin
neighbors.   In the ground state of the striped phase (at any
concentration), we assume that all the spins are contained in
clusters consisting of alternating stripes of spins and vacancies.
Since each spin in such a cluster has two nearest-neighbor spins and
{\em no} second-nearest-neighbor spins, the ground state energy per
spin in this phase is simply
\begin{equation}
E^{stripes} = -J \label{eq:Estripes}.
\end{equation}
$B_c$ is obtained by setting these two energies equal, which gives
$B_c = 0.5J$.   For $B <B_{c}=0.5J$, the ground state is
phase-separated, whereas for $B
> B_c$, the ground state is striped, independent of $p$.

We have verified this prediction numerically by varying $B$ at fixed
$T$ and several values of $p$.  In agreement with eqs.\
(\ref{eq:Eps}) and (\ref{eq:Estripes}), we find that the system, for
any $p$, in the limit of low $T$, always phase separates if $B<B_c$
and forms stripes if $B>B_c$.

%%%%%%%%%%%%%%%%%%%%%%%%%%%%%%%%%%%%%%%%%%%%%%%%%%%%%%%%%%%%%%%%%%%%%%%%%
\subsection{Low-Temperature Helicity Modulus}

For all the low-$T$ configurations shown in
Figs.~(\ref{fig:p0.5_nn2_B0.1})-(\ref{fig:p0.8_nn3_A1.5}), we have
compared the $\gamma(T)$'s obtained from Monte Carlo simulations at
low $T$ to those obtained from eq.\ (\ref{eq:conductivity}). The
configurations are extracted from the snapshots at $T = 0.025$, and
the effective conductances are calculated by numerically solving the
system of linear equations obtained by application of Kirchhoff's
equations to each of the nodes in the network.  To minimize finite
size effects, we used periodic boundary conditions in the direction
perpendicular to that for which we calculated the conductances. The
diagonalization of the resulting matrix was carried out using
Mathematica's built-in function 'Solve', which uses the Gaussian
elimination method~\cite{gentle}.

%DANIEL, need a reference here, plus a few more words saying what you did.
%I am just guessing what you did here.
% DR. STROUD, I added the reference and some more explanation

In Table~\ref{table2}, we show the conductances for these networks,
in both the $x$ and $y$ directions, as well as the corresponding
values of the helicity moduli at $T = 0.025$.
% The values of the conductances at $c = 1$ can be obtained
%analytically, and the corresponding values of the helicity moduli
%are also known (NEED REFERENCE HERE).
% DR. STROUD, when doing the numerical calculation of both helicity and
% conductivity the normalization condition was implicit in the equations, so
% I din't have to use the values corresponding to  c = 1
Evidently, Eq.~(\ref{eq:conductivity}) is well satisfied for the
parameters considered. Where there are discrepancies, we believe
that the source of the error is primarily the Monte Carlo
simulations, since at very low $T$, most of the attempted Monte
Carlo moves are rejected and the phase space sampling may be
insufficient to give an accurate equilibrium average.  We conclude
that our MC simulations are, in general, quite accurately converged,
and also that the mapping proposed in Ref.~\cite{ebner_stroud} is
well obeyed for this rather extensive series of models.

%DANIEL - what are you comparing with what?  In the MC case, do you pick one of
%the low temperature snapshots, and do you calculate the resistance of that snapshot,
%or how do you do this exactly?
% DR. STROUD, yes, I do that exactly.
%
 \begin{table}[hbtp]
   \begin{center}
     \begin{tabular}{||c|c|c||c|c|c|c||}\hline\hline
%       p   &   A    &  B &   $\frac{\gamma_{xx}(0,c)}{\gamma(0,c=1)}$  &  $\frac{g_{xx}(c)}{g(c=1)}$ & $\frac{\gamma_{yy}(0,c)}{\gamma(0,c=1)}$ &  $\frac{g_{yy}(c)}{g(c=1)}$\\ \hline
       p   &   A    &  B &   $\frac{\gamma_{xx}(T=0.025,p)}{\gamma_{xx}(T=0.025,p=1)}$  &  $\frac{g_{e,xx}(p)}{g_{e,xx}(p=1)}$ & $\frac{\gamma_{yy}(T=0.025,p)}{\gamma_{yy}(T=0.025,p=1)}$ &  $\frac{g_{e,yy}(p)}{g_{e,yy}(p=1)}$\\ \hline
       0.5 &  0.0  &  0.1  &    0.50       &   0.48        & -0.01       &    0.00       \\
       0.5 &  0.0  &  1.0  &    0.00       &   0.00        &  0.49       &    0.48       \\
       0.5 &  0.5  &  0.0  &    0.50       &   0.48        &  0.01       &    0.00       \\
       0.5 &  1.5  &  0.0  &    0.00       &   0.00        &  0.00       &    0.00       \\ \hline\hline

       0.8 &   0.0  &  0.1  &    0.62       &   0.58        & 0.67        &    0.66      \\
       0.8 &   0.0  &  1.0  &    0.03       &   0.00        & 0.77        &    0.76       \\
       0.8 &   0.5  &  0.0  &    0.64       &   0.65        & 0.64        &    0.62       \\
       0.8 &   1.5  &  0.0  &    0.26       &   0.17        & 0.36        &    0.48       \\ \hline\hline

     \end{tabular}
     \caption{Comparison of the components $\gamma_{xx}(T=0.025, p)$ and $\gamma_{yy}(T =
       0.025, p)$ for a site- diluted XY-model, as obtained by Monte
       Carlo simulation of Model II, to the effective conductances
       $g_{e,xx}$ and $g_{e,yy}$ of an associated conductance network,
       constructed as described in the text.  $p$ is the spin
       concentration while $A$ ($B$) is the strength of the repulsion
       between first (second) nearest-neighbor spins.}\label{table2}
   \end{center}
 \end{table}
%DANIEL: Note that I have interchanged A and B in the text, so I have also done the same in your
%Table above.  Also, have you normalized any of the above to p = 1?  Why don't you have a
%ratio in the above Table.

%%%%%%%%%%%%%%%%%%%%%%%%%%%%%%%%%%%%%%%%%%%%%%%%%%%%%%%%%%%%%%%%%%%%%%%%%
\section{Discussion}

We have studied a diluted XY model with annealed disorder and an
additional spin-spin repulsion.  We considered two types of
repulsion: (i) a screened Coulomb interaction between spins, with a
finite-separation cutoff, and (ii) a short-range repulsion. In the
first case, we have calculated the types of minimum-energy
configurations found at low $T$, for a variety of parameter choices.
For the second model, we have considered the system at finite and
very low $T$.

For the case of Coulomb repulsion (Model I), we find that,  as the
repulsion strength $C$ increased, the system traverses a series of
ordered phases in a characteristic sequence: first large blobs
(corresponding to a phase-separated state), then horizontal,
vertical and diagonal but straight stripes, then tortuous stripes,
and finally checkerboard-like patterns.  These patterns are
strikingly independent of the model details, such how the Coulomb
interaction is truncated.

For Model II, the low-$T$ spin configuration once again depends on
the relative strength of the attractive XY interaction and the
nearest-neighbor or second-nearest-neighbor repulsion.  The ground
state is always phase-separated for weak enough repulsion, but
becomes either checkerboard or stripe-like for stronger
nearest-neighbor or second-nearest-neighbor repulsion.  For some
concentrations, we see evidence of a {\em phase separation} between
%a checkerboard or striped ordered region, and a region of vacancies.
% DR. STROUD2, I changed the line above to the following, which I assume is what you meant to say
a checkerboard or striped ordered region, and a region with no
vacancies. As $T$ is increased, the spatial ordering of spins,
whether phase-separated, striped, or checkerboard, eventually
disappears in favor of a homogeneous isotropic phase with only
limited short-range order.  The long but tortuous stripe regime
sometimes seen in Model I at low $T$ appears not to occur with Model
II, probably because the repulsion is only short-range.

Corresponding to these $T$-dependent spin distributions in Model II,
we have seen characteristic behavior in the helicity modulus
$\gamma$.  For sufficiently weak repulsion, the system phase
separates at low $T$. In this case, for $p > p_c$, the two diagonal
components of $\gamma$ are approximately equal, and substantially
larger than for quenched disorder. For $p <p_{c}$, phase separation
leads to a non-zero helicity modulus in only one of the two
principal directions. For systems with annealed disorder and strong
second-nearest-neighbors repulsion, the formation of short stripes
leads, at intermediate temperatures, to an increase of the helicity
modulus in both the $x$ and $y$ directions. As $T$ is reduced
further, the stripes become longer but remain randomly oriented,
leading to a reduction of the helicity modulus in both directions.
Finally, as $T \rightarrow 0$, the stripes choose a preferred
direction, and the helicity modulus becomes anisotropic, becoming
large in the direction parallel to the stripes and very small
perpendicular to them.
For strong nearest-neighbor repulsion, the helicity modulus is
always smaller than in systems with quenched disorder.  This
decrease is due to the reduction in number of nearest-neighbor spins
by the repulsive interaction.  In all our calculations, we find
good agreement between the low temperature helicity modulus as
obtained from Monte Carlo and that inferred from the conductance of
an associated conductance network.

%DANIEL, I will try to add some more general discussion and conclusions here.

Finally, we comment on the original motivation for this work, which
was to shed some light on the interplay between inhomogeneities,
stripe and checkerboard order, and superfluid density in the
underdoped cuprate superconductors.  Obviously, the present model is
far too crude to represent all the subtleties of that system.  In
particular, it omits quantum effects arising from the
non-commutativity of number and phase variables\cite{commut}. But
many of the phenomena reported in the cuprates (small superfluid
density, frustrated phase separation, and coexistence of
superconductivity with other types of order, such as checkerboard or
stripe formation) occur in our model. Thus a suitably refined
version of the present model might provide insight into the
interplay between superconductivity and other collective phenomena
in the cuprate superconductors.

\section{Acknowledgments.}  We are grateful for support through the
National Science Foundation, grant DMR04-13395.  Calculations were
carried out using the IA32 Cluster of the Ohio Supercomputer Center.

%%%%%%%%%%%%%%%%%%%%%%%%%%%%%%%%%%%%%%%%%%%%%%%%%%%%%%%%%%%%%%%%%%%%%%%%

%%%%%%%%%%%%%%%%%%%%  Phase diagram %%%%%%%%%%%%%%%%%%%%
%%%%%%%%%%%%%%%%%%%%  Phase diagram %%%%%%%%%%%%%%%%%%%%

%%%%%%%%%%%%%%%%%%%%  Phase diagram %%%%%%%%%%%%%%%%%%%%

%%%%%%%%%%%%%%%%%%%%  Phase diagram %%%%%%%%%%%%%%%%%%%%

%%%%%%%%%%%%%%%%%%%%  Phase diagram %%%%%%%%%%%%%%%%%%%%

%%%%%%%%%%%%%%%%%%%%   

%%%%%%%%%%%%%%%%%%%%    p0.5 strong 2nd nn   %%%%%%%%%%%%%%%%%%%%

%DANIEL, in the above figure, you haven't marked part (a)
%%%%%%%%%%%%%%%%%%%%    p0.5 strong 1st nn   %%%%%%%%%%%%%%%%%%%%

%%%%%%%%%%%%%%%%%%%%   p0.8_L30_nn2    %%%%%%%%%%%%%%%%%%%%

%%%%%%%%%%%%%%%%%%%%   p0.5_L30_nn2    %%%%%%%%%%%%%%%%%%%%

%%%%%%%%%%%%%%%%%%%%   p0.5_L30_nn2    %%%%%%%%%%%%%%%%%%%%

\begin{thebibliography}{99}

%----------------- SELF ORGANIZED SYSTEMS --- EXP

%DANIEL, unless the list is very long, could you replace the et al's by the actual list
% of authors?
%DR. STROUD, done

\bibitem{seul_andelman} Michael Seul and David Andelman, \emph{Science} {\bf 267}, 476 (1995).

\bibitem{huebener} R. P. Huebener, \emph{Magnetic Flux Structures in Superconductors} (Springer-Verlag, 1979.)

\bibitem{kern} Klaus Kern, Horst Niehus, Axel Schatz, Peter
  Zeppenfeld, J\"{u}rgen George, and George Comsa \emph{Phys. Rev.
    Lett.}  {\bf 67}, 855 (1991).

\bibitem{seul_sammon} M. Seul and M. J. Sammon, \emph{Phys. Rev. Lett.}  {\bf 64}, 1903 (1990).

\bibitem{seul_wolfe} M. Seul and R. Wolfe, \emph{Phys. Rev. Lett.}  {\bf 68}, 2460 (1992).

%----------------- STRIPES --- EXP
\bibitem{cheong_aeppli} S.-W. Cheong, G. Aeppli, T. E. Mason, H.
  Mook, S. M. Hayden, P. C. Canfield, Z. Fisk, K. N. Clausen, and J.
  L. Martinez, \emph{Phys. Rev. Lett.}  {\bf 67}, 1791 (1991).
  
\bibitem{yamada} K. Yamada, C. H. Lee, K. Kurahashi, J. Wada, S.
  Wakimoto, S. Ueki, H. Kimura, Y. Endoh., S. Hosoya, G. Shirane, R.
  J. Birgeneau, M. Greven, M. A. Kastner, and Y. J. Kim \emph{Phys.
    Rev. B.}  {\bf 57}, 6165 (1998).

\bibitem{tranquada} J. M. Tranquada, B. J. Sternlieb, J. D. Axe, Y.
  Nakamura, S. Uchida, \emph{Nature} {\bf 375}, 561 (1995).

\bibitem{niemoller} T. Niem\"{o}ller, N. Ichikawa, T. Frello, H.
  H\"{u}nnefeld, N.H. Andersen, S. Uchida, J. R. Schneider, and J. M.
  Tranquada, \emph{Eur. Phys. J.}  {\bf B 12}, 509 (1999).

\bibitem{mook} H. A. Mook, Pengcheng Dai, S. M. Hayden, G. Aeppli, T.
  G. Perring, and F. Dogan, \emph{Nature} {\bf 395}, 580 (1998).

\bibitem{arai} M. Arai, T. Nishijima, Y. Endoh, T. Egami, S.
  Tajima, K. Tomimoto, Y. Shiohara, M. Takahashi, A. Garrett, and
  S. M. Bennington, \emph{Phys. Rev. Lett.}  {\bf 83}, 608 (1999).

\bibitem{pengcheng} Pengcheng Dai, H. A. Mook, R. D. Hunt, F. Dogan,
  \emph{Phys. Rev. B} {\bf 63}, 54525 (2001).

\bibitem{hayden} S. M. Hayden, H. A. Mook, Pengcheng Dai, T. G.
  Perring, and F. Dogan, \emph{Nature} {\bf 429}, 531 (2004).

\bibitem{sun} X. F. Sun, Y. Kurita, T. Suzuki, Seiki Komiya, and
  Yoichi Ando \emph{Phys. Rev. Lett.}  {\bf 92}, 47001 (2004).

%------------------ SELF ORGANIZED SYSTEMS --- THEORY (ISING MODELS)

\bibitem{booth} I. Booth, A. B. MacIsaac, and J. P. Whitehead, and K.
  De'Bell, \emph{Phys. Rev. Lett.}  {\bf 75}, 950 (1995).

\bibitem{stoycheva} Antitsa D. Stoycheva and Sherwin J. Singer, \emph{Phys.
    Rev. Lett.}  {\bf 84}, 4657 (2000).

\bibitem{macisaac} A. B. MacIsaac, J. P. Whitehead, M. C. Robinson,
  and K. De'Bell, \emph{Phys. Rev. B}, {\bf 51}, 16033 (1995).

\bibitem{low_emery} U. L\"{o}w, V. J. Emery, K. Fabricius, and S. A.
  Kivelson, \emph{Phys. Rev. Lett.} {\bf 72}, 1918 (1994).

\bibitem{jamei} Reza Jamei, Steven Kivelson, and Boris Spivak,
  \emph{Phys. Rev. Lett.} {\bf 94}, 056805 (2005).

\bibitem{abanov} A. Abanov, V. Kalatsky, V. L. Pokrovsky, and W. M.
  Saslow, \emph{Phys. Rev. B} {\bf 51}, 1023 (1995).

\bibitem{arlett} J. Arlett, J. P. Whitehead, A. B. MacIsaac, and K.
  De'Bell \emph{Phys. Rev. B} {\bf 54}, 3394 (1996).

\bibitem{hurley} M. M. Hurley and Sherwin J. Singer, \emph{J. Phys
    Chem.}  {\bf 96}, 1938 (1992).

\bibitem{czeck} R. Czech and J. Villain \emph{J. Phys.: Condens.
    Matter} {\bf 1} 619 (1989).

\bibitem{iglesias} J. R. Iglesias, S. Gon\c{c}alves, O. A. Nagel, and
  Miguel Kiwi, \emph{Phys. Rev. B} {\bf 65}, 064447 (2002).

\bibitem{emery_kivelson_tranquada} See V. J. Emery, S. A. Kivelson and
  J.  M. Tranquada, \emph{Proc. Natl. Acad. Sci. USA} {\bf 96}, 8814
  (1999), and references therein.

\bibitem{reichhardt} C. Reichhardt, C. J. Olson Reichhardt, I. Martin,
  and A. R. Bishop, Phys. Rev. Lett. {\bf 90}, 026401 (2003).

%------------------- INHOMOGENEITY IN CUPRATES---EXP
\bibitem{lang_davis} K. M. Lang, V. Madhavan, J. E. Hoffman, E. W.
  Hudson, H. Eisaki, S. Uchida, and J. C. Davis, \emph{Nature} {\bf
    415}, 412 (2002).

\bibitem{pan_davis} S. H. Pan, J. P. O'Neal, R. L. Badzey, C. Chamon,
  H. Ding, J. R. Engelbrecht, Z. Wang, H. Eisaki, S. Uchida, A. K.
  Gupta, K.-W. Ng, E. W. Hudson, K. M. Lang, and J. C. Davis,
  \emph{Nature} {\bf 413}, 282 (2001).

\bibitem{hanaguri_davis} T. Hanaguri, C. Lupien, Y. Kohsaka, D.-H.
  Lee, M. Azuma, M. Takano, H. Takagi, and J. C. Davis, \emph{Nature} {\bf
    430}, 1001 (2004).

%------------------- VARIOUS
\bibitem{stroud_zeng} X. C. Zeng, D. Stroud, and J.S. Chung, \emph{Phys. Rev.
\textbf{B}} {\bf 43}, 3042 (1991).

\bibitem{stroud_shih} W. Y. Shih, C. Ebner, and D. Stroud, \emph{Phys. Rev.
\textbf{B}} {\bf 30}, 134 (1984).

\bibitem{carlson_manousakis} E. W. Carlson, S. A. Kivelson, V. J.
  Emery, and E. Manousakis, \emph{Phys. Rev. Lett.} {\bf 83}, 612
  (1999).

\bibitem{schultka_manousakis} N. Schultka and E. Manousakis, \emph{Phys. Rev.
\textbf{B}} {\bf 49}, 12071 (1994).

\bibitem{moon} K. Moon and S. M. Girvin, \emph{Phys. Rev. Lett.} {\bf
    75}, 1328 (1995).

%-------------------
\bibitem{barkema} M. E. J. Newman and G. T. Barkema, \emph{Monte Carlo
    Methods in Statistical Physics} (Oxford:~Clarendon~Press, 1999.)

\bibitem{ebner_stroud} C. Ebner and D. Stroud, \emph{Phys. Rev.
    \textbf{B}} {\bf 28}, 5053 (1983).

\bibitem{emery} V. J. Emery, E. Fradkin, S. A. Kivelson, and T. C.
Lubensky, Phys.\ Rev.\ Lett.\ {\bf 85}, 2160 (2000).

\bibitem{roddick_stroud} Eric Roddick and David Stroud,
  \emph{Phys. Rev. Lett.}  {\bf 74}, 1430 (1995).

\bibitem{carlson_kivelson} E. W. Carlson, V. J.  Emery, S. A.
  Kivelson, and D. Orgad. The Physics of Superconductors, Vol. 2. Eds.
  K.  H. Bennemann and J. B. Ketterson, Springer-Verlag (2004).

\bibitem{albuquerque} M. Portes de Albuquerque and P. Molho,
  \emph{Journal of Magnetism and Magnetic Materials}, {\bf 113}, 132
  (1992).

\bibitem{sahimi} Muhammad Sahimi, \emph{Applications of Percolation
    Theory} (Taylor and Francis, 1994.)

%******

\bibitem{gentle} J. E. Gentle, \emph{Numerical Linear Algebra for Applications in Statistics} (Springer-Verlag, 1998.)

%\bibitem{chen_cheong} C. H. Chen, S-W. Cheong, and A. S. Cooper,
%  \emph{Phys. Rev. Lett.}  {\bf 71}, 2461 (1993).

\bibitem{commut} See, e. g., S. L. Sondhi, S. M. Girvin, J. P. Carini, and D. Shahar, Rev.\ Mod.\
Phys.\ {\bf 69}, 315 (1997) and references therein.

\end{thebibliography}
\end{document}